\documentclass[final,sort&compress]{elsarticle}

\usepackage{graphicx,amsmath,amssymb,subfigure,bbm,epstopdf,soul}
\usepackage{algpseudocode,algorithm,color}
\usepackage{soul}
\usepackage{caption,url}
\usepackage[makeroom]{cancel}

\newcommand{\Sm}{\mathbf{S}}
\newcommand{\Vm}{\mathbf{V}}
\newcommand{\Um}{\mathbf{U}}
\newcommand{\Dm}{\mathbf{D}}
\newcommand{\T}{\mathbf{T}}

\newcommand{\Jfunc}{\mathcal J }
\newcommand{\Model}{\mathcal M}
\newcommand{\M}{\mathbf{M}}

\newcommand{\Hobs}{\mathcal{H}}
\newcommand{\HH}{\mathbf H}

\newcommand{\A}{\mathbf{A}}
\newcommand{\B}{\mathbf{B}}
\newcommand{\C}{\mathbf{C}}
\newcommand{\R}{\mathbf{R}}

\newcommand{\Q}{\mathbf{Q}}

\newcommand{\V}{ \mathbf{V} }
\newcommand{\W}{ \mathbf{W} }

\newcommand{\x}{   \mathbf{x} }
\newcommand{\xb}{ \mathbf{x}^{\rm b} }
\newcommand{\xa}{ \mathbf{x}^{\rm a} }

\newcommand{\xv}{ \mathbf{x}^{\rm v} }
\newcommand{\y}{ \mathbf{y} }

\renewcommand{\u}{ \mathbf{u} }

\renewcommand{\H}{\mathcal{H}}

\newcommand{\Y}{\mathcal{Y}}

\graphicspath{ {./Figures/} }

\begin{document}

\thispagestyle{empty}
\setcounter{page}{0}

\begin{Huge}
\begin{center}
Computational Science Laboratory Technical Report CSL-TR-3-2013 \\
\today
\end{center}
\end{Huge}
\vfil
\begin{huge}
\begin{center}
Alexandru Cioaca, \\ Adrian Sandu, and Eric de Sturler
\end{center}
\end{huge}

\vfil
\begin{huge}
\begin{it}
\begin{center}
Efficient methods for computing observation impact \\
in 4D-Var data assimilation
\end{center}
\end{it}
\end{huge}
\vfil

\begin{large}
\begin{center}
Computational Science Laboratory \\
Computer Science Department \\
Virginia Polytechnic Institute and State University \\
Blacksburg, VA 24060 \\
Phone: (540)-231-2193 \\
Fax: (540)-231-6075 \\ 
Email: \url{sandu@cs.vt.edu} \\
Web: \url{http://csl.cs.vt.edu}
\end{center}
\end{large}

\vspace*{1cm}

\begin{tabular}{ccc}
\includegraphics[width=2.5in]{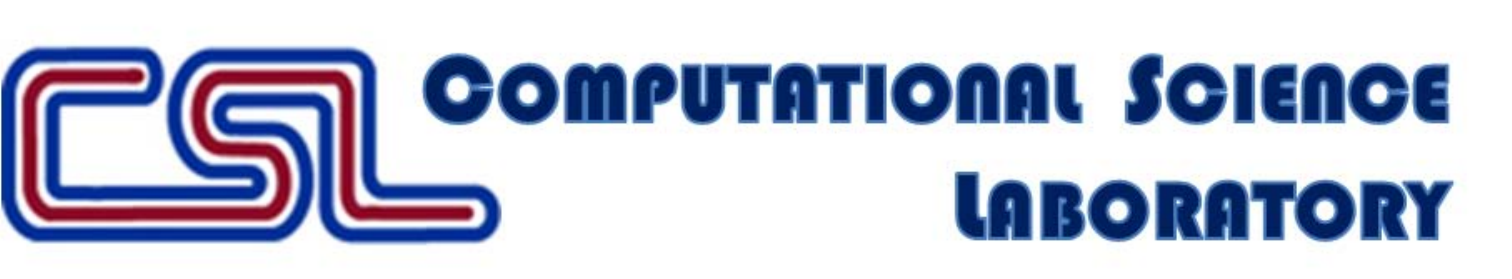}
&\hspace{2.5in}&
\includegraphics[width=2.5in]{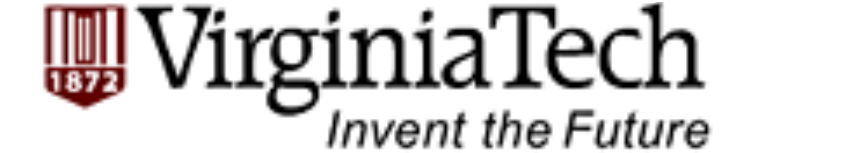} \\
{\bf\em Innovative Computational Solutions} &&\\
\end{tabular}

\newpage

\title{Low-rank Approximations \\ for Computing Observation Impact \\ in 4D-Var Data Assimilation}

\date{\today}

\author{Alexandru Cioaca}
\author{Adrian Sandu}


\begin{abstract}

We present an efficient computational framework to quantify the impact of individual observations in four dimensional variational data assimilation.
The proposed methodology uses first and second order adjoint sensitivity analysis, together with
matrix-free algorithms to obtain low-rank approximations of observation impact matrix. 
We illustrate the application of this methodology to important applications such as data pruning and the identification of faulty sensors
for a two dimensional shallow water test system.
 
\end{abstract}

\maketitle

\newpage

\tableofcontents

\newpage
\setcounter{page}{1}

\section{Introduction}

Data assimilation is a dynamic-data driven application that integrates information from physical observations with numerical model predictions. 
This paper describes a systematic approach to quantify the contribution of each observation data point in improving the model state estimates.
The focus is on the four-dimensional variational (4D-Var) data assimilation approach, which fits a model trajectory against time-distributed observations,
to obtain maximum likelihood estimates of the initial state and model parameters. 

Computing the ``observation impact'' (OI) means quantifying the individual (or group) contributions of ``assimilated'' data points to 
reduce the uncertainty in state or model parameters. OI provides a numerical measure to distinguish between crucial and redundant information,
and benefits applications such as big data pruning, error detection/correction, 
intelligent sensor placement, and other decision-making processes.

There are several approaches available in the scientific literature to compute OI. 
In early studies OI was associated with the energy of perturbations along the dominant directions of maximum error growth \cite{Palmer_1999, Sandu_HSV}. 
Other approaches study observation impact using metrics from information theory \cite{ZupInfoTheory,Singh_2012}, 
statistical design \cite{Berliner_1999}, and robust control \cite{TELLUSA17133}. 
More recently, observation impact has been assessed through adjoint-sensitivity analysis,
as a counterpart measure to observation sensitivity \cite{gelaro2007examination,gelaro2009examination,TELA:TELA349}.
The current state-of-the-art uses the second-order sensitivity equations of 4D-Var system\cite{LeDimet_1997,Daescu_2008,Daescu_2010},
a powerful approach followed by the present research study.
The efficient implementation of this methodology is hindered by several computational limitations
related to the calculation of second-order derivative approximations.

This work develops a systematic and efficient approach to compute OI in a sensitivity analysis framework.
Observation impact is formalized as a sensitivity matrix which maps changes in observation space 
to associated changes in solution space. This matrix is usually not available in explicit (full) form. 
Our computational approach makes smart use of tangent linear and adjoint models \cite{cacuci1981sensitivity,wang1992second,sandu2008discrete}
to propagate first and second order sensitivities through the data assimilation system, and obtain products of the OI matrix with user defined vectors.
Matrix-free linear algebra tools are necessary for this. 
Eigenvalue and singular value decompositions are used to obtain low-rank approximations that capture the main features of observation impact,
while significantly reducing the computational burden. 
As of recently, low-rank approximations of matrices have become very popular in image processing \cite{ye2005svdvision}, 
information retrieval \cite{berry1996low}, and machine learning \cite{dhillon2001concept} to extract correlations and 
remove noise from data. Two alternative ways to compute a low-rank approximation for observation impact are given,
one of serial nature and one highly parallel. 
Several applications of OI are exemplified using the two dimensional shallow water equations test problem.

The remaining part of the paper is structured as follows. 
Section \ref{sec:4dvar} reviews the formulation of 4D-Var data assimilation.
Section \ref{sec:oim} reviews the 4D-Var sensitivity equations necessary for
the observation impact matrix (following the original derivation \cite{Daescu_2008}). 
We discuss the structural and functional properties of the impact matrix and the required numerical tools for evaluating its action.
Section \ref{sec:lowrank} introduces low-rank approximations of the observation impact matrix. 
Numerical experiments to exemplify and validate the observation impact matrix are presented in Section \ref{sec:results}. 
Conclusions and future directions of research are outlined in Section \ref{sec:concl}.

\section{4D-Var Data Assimilation}\label{sec:4dvar}

\subsection{Formulation}

Data assimilation is the process by which model predictions are constrained with real measurements \cite{Daley_1991,Kalnay_2002}. 
It combines three sources of information: an a priori estimate of the initial state of the system (``background''), 
knowledge of the physical laws governing the behavior of the system (captured by a numerical model), and measurements
of the real system (``observations''). Data assimilation solves the inverse problem for improving 
estimates of model states, initial and boundary conditions, or various system parameters.

Four-dimensional variational (4D-Var) assimilation is formulated as a PDE-constrained nonlinear optimization.
The improved initial state $\xa_0$ (``analysis'') is obtained by minimizing the following cost function:
 \begin{subequations}
 \begin{eqnarray}
  \label{eqn:fdvar-costfun}
  \Jfunc(\x_0) &=& \frac{1}{2} \left( \x_0 - \xb_0 \right)^T \cdot \B_0^{-1} \cdot ( \x_0 - \xb_0 )   \\
             & & \, + \frac{1}{2} \sum_{k=1}^{N} \left( \Hobs_k (\x_k) - \y_k \right)^T \cdot \R_k^{-1} \cdot \left( \Hobs_k (\x_k) - \y_k \right)\,, 
  \nonumber \\
 \label{eqn:fdvar-optimization}  
 \xa_0 &=& \arg\min_{\x_0} \Jfunc(\x_0) \quad \textnormal{subject to }\x_k = \Model_{t_0 \rightarrow t_k} (\x_0)\,.
 \end{eqnarray}
 \end{subequations}
Here $\Model$ denotes the numerical model used to evolve the initial state vector $\x_0$ in time. 
$\Hobs_k$ denotes the observation operator at assimilation time $t_k$, and maps the model state 
$\x_k \approx \x(t_k)$ to the observation space. 
$\R_k$ is the observation error covariance matrix. 
A practical assumption is that observations are independent, hence uncorrelated, which reduces $\R_k$ to a diagonal
matrix containing observation variances. 
The covariance matrices $\B_0$ and $\R_k$ are predefined, and the
 quality of their approximation influences the resulting analysis.

The first term of the 4D-Var cost functional \eqref{eqn:fdvar-costfun} quantifies the mismatch between the initial solution ($\x_0$) 
and the background state ($\xb_0$) at the initial time ($t_0$). 
This mismatch is computed in a least-squares sense, scaled by the inverse background error covariance matrix $\B_0$. 
The second term measures the mismatch between the model trajectory (initialized from $\x_0$) and observations $\y_k$
taken at times $t_k$, $k=0,\dots,N$ scaled by the inverse observation error covariances $\R_k$.
Through $\B_0$ and $\R_k$ the inverse problem takes into account the uncertainty in the data and model predictions. 
4D-Var can be viewed as a method of Bayesian inference, which computes the maximum likelihood initial solution 
conditioned by the observations.

When assimilating observations only at the initial time $t_0$ the method is known as three-dimensional variational 
(3D-Var), as the additional time dimension is not present. 

\subsection{Computational aspects}

The minimization problem \eqref{eqn:fdvar-optimization} is computed numerically using
gradient-based iterative nonlinear solvers such as
quasi-Newton, nonlinear conjugate gradients, and truncated Newton methods. 

The iterative solver starts from an initial guess and advances to a minimizer,
along directions of descent computed using the gradients of $\Jfunc$. 
The descent directions depend on all three available sources of information;
intermediate iterations
can be viewed as the result of partially assimilating information given by observations.

The iterative nature of the solution approach has several important consequences.
The resulting analysis is not the global minimizer, but rather is a local one.
When the computational cost per iteration is high, the available
computational resources constrain the number of iterations that can be performed. 
Thus in practice analyses are almost always suboptimal.

We will call the numerical model $\mathcal{M}$ (\ref{eqn:fdvar-optimization}) the {\em forward model} ({\sc fwd}) or forecast model.
Applying gradient-based methods to minimize (\ref{eqn:fdvar-optimization}) requires the derivatives of $\Jfunc$ with respect to the initial model states.
This can be achieved using adjoint modeling \cite{cacuci1981sensitivity,wang1992second},
a methodology that has been successfully implemented in optimization, sensitivity analysis and uncertainty quantification
\cite{sandu2008discrete,SanduADJ_2005,Cioaca_2011}.
The adjoint models ({\sc adj}) can be based either on the linearization of the original differential equations (``continuous'' adjoint approach),
or the linearization of the numerical method (``discrete'' adjoint approach).
A convenient approach to generate discrete adjoint models is automatic differentiation  \cite{griewank1989automatic},
which takes as input the source code of the {\sc fwd} model, performs line by line differentiation, and returns the source code of the adjoint models.

\section{Observation Impact}\label{sec:oim}

\subsection{4D-Var sensitivity to observations}\label{sec:sensan}

We now establish a  relation between the analysis $\xa_0$ and the observations $\y_k$.
This is done using sensitivity analysis, and expresses how slight changes
in the observational data translate into changes in the resulting analysis.
This section follows the 4D-Var sensitivity approach of Daescu \cite{Daescu_2008}.

Consider the problem of finding a vector $\x = (x_1, x_2, ..., x_n)^T \in \mathbb{R}^n$ 
that minimizes the twice continuously differentiable cost function:
\begin{equation}
 \xa = \arg\min_{\x} \Jfunc(\x,\u) \,.
\end{equation}
The function depends on the state vector $\x$ and parameter vector $\u \in \mathbb{R}^m$. 
For any value $\bar{\u}$ of the parameter, the solution
obeys the first order optimality condition
\begin{equation}
 \nabla_\x \, \Jfunc(\xa,\bar{\u}) = 0 \,.
\label{eqn:fooc}
\end{equation}

Assume that the Hessian of the function is positive definite at the optimum, $\nabla_{\x,\x}^2 \Jfunc(\xa,\bar{\u})>0$.
The implicit function theorem applied to \eqref{eqn:fooc} guarantees there exists a vicinity of $\bar{\u}$ where the optimal solution is a smooth function
of the parameters, $\xa = \xa(\u)$, and has the sensitivity
\[
 \nabla_\u\, \xa(\u) = -\nabla_{\u,\x}^2 \Jfunc (\u, \xa) \cdot \left(\nabla_{\x,\x}^2 \Jfunc(\u, \xa)\right) ^{-1}\,. 
\]
We apply this procedure to the 4D-Var cost function \eqref{eqn:fdvar-costfun} with the parameters $\u$ being the observational data $\y_k$. The first-order necessary condition reads:
\begin{eqnarray}
\nabla_{\x_0}\, \Jfunc(\xa_0) = \B_0^{-1} \left(\xa - \xb\right) + \sum_{k=1}^{N}  \M_{0,k}^T \HH_k^T \R_k^{-1} \left( \Hobs_k(\x_k) - \y_k \right) = 0 \,,
\label{eqn:fdvar-fooc}
\end{eqnarray}
where $\M_{0,k} = \Model_{t_0 \rightarrow t_k}'(\x_0)$ is the tangent linear propagator associated with the 
numerical model $\Model$, and $\HH_k=\Hobs_k'(\x_k)$ is the linearized observation operator at time $t_k$. 
Differentiating (\ref{eqn:fdvar-fooc}) with respect to the observations $\y_k$ 
\[
\nabla_{\y_k, \x_0}^2 \, \Jfunc(\xa_0) = -\R_k \, \HH_k\, \M_{0,k}
\]
leads to the following expression for the sensitivity of the analysis to observations:
\begin{eqnarray}
\nabla_{\y_k}\, \xa_0 &=& \left( \frac{\partial \xa_0}{\partial\y_k} \right)^T = \R_k^{-1}\, \HH_k\, \M_{0,k}\, \A_0 \,,
\label{eqn:oi-matrix} \\
\A_0 &=&  \left(\nabla_{\x_0,\x_0}^2 \Jfunc(\xa_0)\right) ^{-1}\,.
\label{eqn:inverse-hessian} 
\end{eqnarray}

Consider a verification cost functional $\Psi : \mathbb{R}^n \rightarrow \mathbb{R}$
that measures the discrepancy between the analysis and a verification solution $\xv_0$:
\begin{align}
\label{eqn:ST0_psifun}
 \Psi ( \xa_0 ) = \frac{1}{2}(\xa_0 - \xv_0)^T \, \C \, (\xa_0 - \xv_0)\,.
\end{align}
Here the verification solution $\xv_0$ is also defined at $t_0$. The matrix $\C$ is a weighting matrix corresponding to a particular norm
or restricting the verification to a subdomain of the solution space. 
Using chain-rule differentiation and \eqref{eqn:oi-matrix}, 
the \emph{sensitivity to observations} (the gradient of $\Psi$ with respect to observations $\y_k$) is:
\begin{align}
\label{eqn:ST0_psigrad}
\nabla_{\y_k} \Psi ( \xa_0 ) \,=&\, \nabla_{\y_k}\, \xa_0 \, \cdot \, \nabla_{\xa_0} \Psi ( \xa_0 ) \\
\nonumber                            \,=&\, \R_k^{-1}\, \HH_k\, \M_{0,k}\, \A_0 \, \C \, (\xa_0 - \xv_0)\,.
\end{align}
The sensitivity \eqref{eqn:ST0_psigrad} can be computed via the following steps:
\begin{itemize}
\item The sensitivity of the verification function with respect to the analysis is 
 \[
  \nabla_{\xa_0} \Psi(\xa_0) = \C \, (\xa_0 - \xv_0)\,.
 \]
\item A ``supersensitivity'' is obtained through solving a linear system with
the matrix the 4D-Var Hessian {eqn:inverse-hessian} 
 \begin{align}
  \mu = \A_0 \cdot \nabla_{\xa_0} \Psi(\xa_0)\,.
  \label{eqn:ST1_supersens}
 \end{align}
This step dominates the computational cost of the sensitivity calculation.
\item Finally, the verification sensitivity to observations $\y_k$ is the
supersensitivity vector propagated to time $t_k$ through the tangent linear model, 
mapped through the observation selection operator, and scaled with the inverse error covariance:
 \begin{align}
  \nabla_{\y_k} \Psi(\xa_0) = \R_k^{-1}\, \HH_k\, \M_{0,k}\, \cdot \mu\,.
  \label{eqn:ST2_obssens}
 \end{align}
\end{itemize}
The sensitivities of $\Psi$ to the background $\xb_0$ and to the error covariance matrices $\B$ and $\R$
can be derived in a similar fashion \cite{Daescu_2008}.

\subsection{Observation impact matrix}\label{sec:obsimp}

\subsubsection{Main features}

Equation (\ref{eqn:oi-matrix}) defines a matrix whose elements are
the sensitivities of each component of the analysis vector $\xa_0$ to each component 
of the observation vector $\y_k$ assimilated at time $t_k$. 
The {\em observation impact matrix} collects the sensitivities to all observations as follows:
\begin{equation}
\label{eqn:oimatext}
\T
= 
\left(\frac{\partial \xa}{\partial \y}\right)^T
=
\begin{pmatrix}
 \nabla_{\y_1} \xa_0 \\[0.5em]
 \nabla_{\y_2} \xa_0 \\[0.5em]
 \vdots 	             \\[0.5em]
 \nabla_{\y_N} \xa_0 \\[0.5em]
\end{pmatrix}
=
\begin{pmatrix}
 \R_1^{-1}\, \HH_1\, \M_{0,1} \, \\[0.5em]
 \R_2^{-1}\, \HH_2\, \M_{0,2} \, \\[0.5em]
 \vdots   \\[0.5em]
 \R_N^{-1}\, \HH_N\, \M_{0,N} \, \\[0.5em]
\end{pmatrix}\,
\A_0\,. 
\end{equation}

Each of the $n$ columns of $\T$ represents the sensitivities of one particular model state to all observations. 
For real applications $n \sim 10^7 - 10^{10}$. 
Each row of $\T$ contains the sensitivities of each state to one particular observation.
In typical data assimilation applications the number of observations (rows) 
is two-three orders of magnitude smaller than the number of model states (columns). 

We now seek to understand the structure of the observation impact matrix (\ref{eqn:oimatext}), 
which is the transpose of the sensitivity matrix $\partial \xa/\partial \y$.
The impact matrix is the product of submatrices of type $\R_k^{-1}\, \HH_k\, \M_{0,k}$, with
the inverse of the 4D-Var Hessian \eqref{eqn:inverse-hessian}. 
The Hessian is symmetric and positive-definite when evaluated at the minimum, but can
lose positive-definiteness when evaluated at an inexact analysis, such as when the minimization of 4D-Var
was incomplete. The inverse of the 4D-Var Hessian at the minimum approximates the analysis error covariance \cite{gejadze2008analysis}.
This aposteriori error covariance characterizes the 4D-Var process and quantifies the uncertainty reduction due to assimilating the observations. 
Each column of this matrix represents the error covariance corresponding to a certain model state. 
The tangent linear model $\M_{0,k}$ propagates this error covariance to the observational time $t_k$.
The propagated perturbations are mapped to observation space through $\H_k$, and then scaled with $\R_k^{-1}$. 
The inverse observation error covariance can be interpreted as a measure of trust assigned to each data point.
Large values of the result are associated to points in space and time where better measurements can benefit the assimilation process. 
This metric can be assessed not only at times when we have observations available, but also
at intermediate times, and can prove to be useful for deploying adaptive observations.

Small changes in observations $\Delta\y$ lead to a change in the analysis $\Delta \xa_0$ which, to first order, can be calculated using the observation impact matrix \eqref{eqn:oimatext} as
\begin{equation}
\label{eqn:T-mapping}
\Delta \xa_0 = \T^T \cdot \Delta \y\,.
\end{equation}
We call the change in analysis $\Delta \xa_0$ the {\em observation impact}.  It is computed via the following steps:
\begin{itemize}
\item Each observation change $\Delta\y_k$ is scaled and pulled back to the initial time via the adjoint model:
\[
  \overline{\Delta \y}_k =\M_{0,k}^T\, \HH_k^T\,  \R_k^{-1}\, \cdot \Delta \y_k\,.
\]
\item The aggregated contribution of all observation changes is
\[
  \overline{\Delta \y} = \sum_{k=0}^N\, \overline{\Delta \y}_k\,,
\]
and in practice can be computed via {\em a single} adjoint run.
\item The observation impact is obtained by solving a linear system whose matrix is the 4D-Var cost function Hessian
\[
\Delta \xa_0 = \A_0 \cdot \overline{\Delta \y} \,.
\]
\end{itemize}

\subsubsection{Computational ingredients}

For real models the matrices that appear in \eqref{eqn:oimatext} are very large. 
The calculations of observation impact rely on matrix-vector products. 
The linear operator $\M_{0,k}$ is applied by running the tangent linear model from $t_0$ to $t_k$. 
Application of the linearized observation mapping operator $\H_k$ and scaling by 
$\R_k^{-1}$ can be performed directly as they are relatively inexpensive.

The inverse Hessian has been successfully used in practical applications to compute the supersensitivity.
This was done by solving iteratively a linear system for the supersensitivity, the system matrix being the 4D-Var Hessian.
Iterative solvers of Krylov-type require only Hessian-vector products. 
These products can be computed by running the second-order adjoint model.
Furthermore, the linear solvers can be preconditioned for faster convergence and several matrix-free methods are readily available \cite{Cioaca_2012}.
When the second-order adjoint is not available, several approximations can be used as follows.

\begin{enumerate}

\item The finite difference of gradients
\[
\nabla_{\x_0,\x_0}^2 \mathcal \Jfunc(\xa_0) \cdot \u \approx \frac{\nabla_{\x_0} \mathcal \Jfunc(\xa_0 + \epsilon \cdot \u)^T - \nabla_{\x_0} \mathcal \Jfunc(\xa_0)^T}{\epsilon}\,.
\]
requires two first-order adjoint runs, which can be performed in parallel. The accuracy of this approximation is typically low.
 
\item The Gauss-Newton approximation of the Hessian
 
\[
\nabla_{\x_0,\x_0}^2 \mathcal \Jfunc(\xa_0) \cdot \u \approx \B_0^{-1}  \cdot \u + \sum_{k=1}^{N}  \M_{0,k}^T \HH_k^T \R_k^{-1}  \HH_k\, \M_{0,k}\cdot \u
\]
discards second order terms which contain the observation residual, so this approximation is independent of the observation data. It requires one tangent-linear model run 
followed by one first-order adjoint run.
 
\item Limited memory quasi-Newton approximations are based on the sequence of solutions and gradients generated during the numerical optimization procedure.
An example of such an approximation is L-BFGS \cite{zhu1997algorithm}.
  
\end{enumerate}

\section{Efficient Computation of Observation Sensitivity and Impact}\label{sec:lowrank}

The computation of observation sensitivity or impact is a non-trivial task since for practical problems 
of interest it is not feasible to build the full matrices.
We seek to develop computationally inexpensive approximations of the impact matrix that capture 
the most important features, and whose accuracy can be improved with an increased computational effort. 
Our approach is based on the Singular Value Decomposition (SVD), a powerful tool to generate low-rank approximations of large matrices.
We present two algorithms, one iterative (inherently serial), and one ensembled-based (inherently parallel).

Consider the SVD of \eqref{eqn:oimatext} and the corresponding low-rank approximations
\begin{eqnarray*}
\T = \Um\,\Sm\,\Vm^T\,, \quad \T_{(p)} = \Um_{(p)}\,\Sm_{(p)}\,\Vm_{(p)}\,,
\end{eqnarray*}
where $\Um$ and $\Vm$ are orthogonal, $\Sm$ is diagonal,  $\Um_{(p)}$ and $\Vm_{(p)}$ are the right and the left singular vectors 
associated with the largest $p$ singular values, and $\Sm_{(p)}$ has these dominant singular values on the diagonal formed.

$\T_{(p)}$ has the smallest ''reconstruction error'' in both Frobenius $\|\T-\T_{(p)}\|_F$ and $L^2$ norms $\|\T-\T_{(p)}\|_2$ among all the rank $p$ approximations of $\T$. 
The accuracy increases as more dominant singular modes are added.
The cut-off threshold is particular to the problem under study and can be determined from the singular value spectrum decay. 

\subsection{An iterative (serial) approach for matrix-free low-rank approximations with SVD}

An ideal iterative SVD algorithm for our problem uses one matrix-vector product per iteration, 
and reveals one new singular pair with each iteration.
The singular vectors are discovered in decreasing order of the magnitude of their singular values.
Thus, running the algorithm for $p$ iterations would generate the leading $p$ singular values and their associated singular vectors.

There are no classic algorithms to compute iteratively the SVD of matrix available only in operator form.
We change our problem to computing the leading eigenvectors of the product between the observation impact matrix and its transpose, $\T^*\,\T = \Um \Sm^2 \Um^*$, where
\begin{eqnarray}
\label{eqn:tstar-t}
 \T^*\,\T &=& \A_0^* \, \sum_{k=1}^N \overline{\M}_{0,k}^* \overline{\M}_{0,k} \,  \A_0 =  \sum_{k=1}^N \T_k^* \T_k\,, \\
 \nonumber
\overline{\M}_{0,k} &=&  \R_k^{-1}\, \HH_k\, \M_{0,k}\,, \quad
\T_k = \overline{\M}_{0,k}  \, \A_0\,.
\end{eqnarray}
This problem is solved using Krylov-based approaches (Lanczos \cite{lanczos1950iteration}, Arnoldi \cite{arnoldi1951principle}), e.g.,
by the Jacobi-Davidson algorithm \cite{sleijpen2000jacobi}, available in the JDQZ software library \cite{fokkema1999short}.

Our algorithm computes a low-rank approximation of $\A_0$, the inverse 4D-Var Hessian.
An iterative procedure can be used to compute the smallest eigenvalues and the corresponding eigenvectors of the Hessian matrix,
which are the dominant eigenpairs of $\A_0$.
The low-rank approximation of the inverse Hessian reads:
\begin{eqnarray}
 \label{eqn:svdhess}
 \A_0 = \left(\Vm\,\Dm\,\Vm^*\right)^{-1} = \Vm\,\Dm^{-1}\,\Vm^* \approx \left(\Vm_{(p)}\,\Dm_{(p)}^{-1}\,\Vm_{(p)}^*\right)\,.
\end{eqnarray}

We replace the inverse Hessian in \eqref{eqn:tstar-t} with its low-rank approximation: 
\begin{eqnarray}
 \T_k^*\,\T_k &\approx & \left(\Vm_{(p)} \,\Dm_{(p)}^{-1} \, \Vm_{(p)}^*\right)\,\overline{\M}_{0,k}^*\,\overline{\M}_{0,k}\,\left(\Vm_{(p)}\,\Dm_{(p)}^{-1}\,\Vm_{(p)}^*\right) \nonumber \\
           &=& \Vm_{(p)}\,\Dm_{(p)}^{-1}\,\W_k^*\,\W_k\,\Dm_{(p)}^{-1}\,\Vm_{(p)}^*\,, \nonumber \\
\W_k &=& \overline{\M}_{0,k}\,\Vm_{(p)}   \,.       \label{eqn:tlmeig} 
\end{eqnarray}

The columns of $\W_k$ are the Hessian eigenvectors $\Vm_{(p)}$ propagated forward and scaled by the tangent linear model $\overline{\M}_{0,k}$. 
The $p$ tangent linear models can be performed in parallel.
This allows us to approximate \eqref{eqn:tstar-t} as:
\begin{eqnarray*}
 \T^*\,\T &\approx& \V_{(p)}\,\Dm_{(p)}^{-1} \, \W \, \Dm_{(p)}^{-1}\,\V_{(p)}^*\,, \\
 \W &=& \sum_{k=1}^N  \W_k^*\,\W_k\,.
\end{eqnarray*}

With the eigendecomposition:
\begin{eqnarray}
 \Dm_{(p)}^{-1}\,\W\,\Dm_{(p)}^{-1} = \Vm_{\rm red} \Dm_{\rm red} \Vm_{\rm red}^*\,, \label{eqn:redsvd}
\end{eqnarray}
the approximation of \eqref{eqn:tstar-t} becomes
\begin{eqnarray}
 \T^*\,\T \approx \left( \Vm_{(p)}\,\Vm_{\rm red} \right) \,\Dm_{\rm red}\,\left( \Vm_{(p)}\,\Vm_{\rm red} \right)^*\,. \label{eqn:obsimplowrank}
\end{eqnarray}
This represents a truncated singular vector decomposition of $\T$, with $\Dm_{\rm red}$ the matrix of dominant singular values, and 
$\Vm_p\,\Vm_{\rm red}$ the matrix of left singular vectors.

\begin{algorithm}[h]
\caption{\label{alg:iterative}Iterative algorithm for low-rank approximations}
\begin{algorithmic}[1]
\State Solve iteratively the eigenvalue problem for the 4D-Var Hessian (\ref{eqn:svdhess})
\State Map newly generated eigenvectors through the tangent linear model (\ref{eqn:tlmeig})
\State Compute the truncated SVD of the resulting matrix (\ref{eqn:redsvd})
\State Project the left singular vectors onto the eigenvector base of the 4D-Var Hessian and build the low-rank approximation of $\T$ (\ref{eqn:obsimplowrank})
\end{algorithmic}
\end{algorithm}

Algorithm \ref{alg:iterative} summarizes the main computational steps. The computational cost is
dominated by the first step, where the expensive second-order adjoint model is run repeatedly to generate the 
Hessian eigenpairs.
The iterative approach is suited for applications that benefit from an iterative improvement of the low-rank approximation. 
The methodology can be applied to any data assimilation system for which first and second order adjoint models are available.

\subsection{An ensemble-based (parallel) approach for matrix-free low-rank approximations}

This approach uses a ``randomized SVD'' algorithm \cite{liberty2007randomized} to compute the Moore-Penrose pseudoinverse \cite{lewis1968pseudoinverss}
of the Hessian. The 4D-Var Hessian matrix $\A_0^{-1}$ is available only in operator form, i.e., only matrix vector products can be evaluated by running the second order adjoint. 
The randomized algorithm is as follows:
\begin{algorithmic}[1]
\State Draw $p$ random vectors and form a matrix $\Omega$.
\State Compute the product $\Y = \A_0^{-1}\Omega$ using Hessian-vector multiplications, i.e., running the second order order adjoint model for each column.
\State Construct the QR decomposition $\Y = \Q\R$.
\end{algorithmic}
Each of the above steps can be performed in parallel.

The columns of $\Q$ form an orthonormal basis for the range of $\Y$.
Randomized SVD uses a series of algebraic manipulations, starting from the observation that $\Q$ is also the
orthonormal factor in the QR decomposition of $\A_0^{-1}$:
\begin{equation}
\label{eqn:aqrb}
\A_0^{-1} =  \Q \, \B \,, \quad
\B = \Q^* \, \A_0^{-1}  \,, \quad
\B^* = \A_0^{-1} \, \Q\,.
\end{equation}
Next, compute an SVD of $\B$:
\begin{eqnarray}
\B = \Um_{\B} \, \Sigma_{\B} \, \V_{\B}^* \label{eqn:svdb}
\end{eqnarray}
and replace (\ref{eqn:svdb}) in (\ref{eqn:aqrb}) to obtain the SVD of $\A_0^{-1}$:
\begin{eqnarray}
\A_0^{-1} &= \Q \, \Um_{\B} \, \Sigma_{\B} \, \V_{\B}^* = \Um_{\A} \, \Sigma_{\B} \, \V_{\B}^*\,. \label{eqn:svda}
\end{eqnarray}

The left singular vectors of $\A$ represent the projections of the left singular vectors of $\B$ onto the columns of $\Q$. 
The singular values and right singular vectors of $\A$ are the same as those of $\B$. 
The pseudoinverse of the 4D-Var Hessian $\A_0^{-1}$ reads:
\begin{equation}
\A_0^+ \approx 
\Vm_{\B} \, \Sigma_{\B}^+ \, \Um_{\A}^* \,.
\label{eqn:pseudo}
\end{equation}

The observation impact matrix is approximated using the tangent linear model $\overline{\M}_{0,k}$ and the pseudoinverse $\A_0^+$:
\begin{eqnarray}
 \T \approx \sum_{k=1}^N \overline{\M}_{0,k} \, \A_0^+\,. \label{eqn:obsimppar}
\end{eqnarray}
The computational flow is summarized in Algorithm \ref{alg:parallel}.

\begin{algorithm}[h]
\caption{Sampling algorithm for low-rank approximations\label{alg:parallel}}
\begin{algorithmic}[1]
\item Build the matrix $\B$, through parallel second-adjoint runs (\ref{eqn:aqrb})
\item Compute a full SVD of $\B$  (\ref{eqn:svdb})
\item Project the left singular vectors of $\B$ in $\Q$ and form the SVD of $\A_0^{-1}$  (\ref{eqn:svda})
\item Compute the Hessian pseudoinverse $\A_0^+$ (\ref{eqn:pseudo})
\item Build the impact matrix $\T$ through parallel tangent linear runs (\ref{eqn:obsimppar})
\end{algorithmic}
\end{algorithm}

Computing the rows of $\B$ \eqref{eqn:aqrb} is done as matrix-vector products through second-order adjoint runs that can be performed in parallel.
The last step, which propagates the components of the pseudoinverse through the linearization of the model,
is achieved by multiple tangent linear model runs in parallel. 
The tangent model results are checkpointed at each of the observation times,
so that only one run across the entire time horizon is necessary for each input vector.

\section{Applications}\label{sec:results}

We illustrate several applications of the observation impact matrix in 4D-Var using the two-dimensional shallow water equations.
We first describe the system and its numerical discretization, then 
present the 4D-Var implementation and the experimental setting for data assimilation.
Observation sensitivity is computed both in full and using low-rank approximation to assess how well the latter captures the essential features.
We apply this analysis to three applications, namely, detecting the change in impact from perfect data to noisy data, 
pruning the least important observations, and detecting faulty sensors.

\subsection{Test problem: shallow water equations}

The two-dimensional shallow-water equations (2D {\sc swe}) \cite{CPA:CPA3160210103} approximate the movement of a thin layer of fluid inside a basin:
\begin{eqnarray}
 \frac{\partial}{\partial t} h + \frac{\partial}{\partial x} (uh) + \frac{\partial}{\partial y} (vh) &=& 0 \nonumber \\
 \frac{\partial}{\partial t} (uh) + \frac{\partial}{\partial x} \left(u^2 h + \frac{1}{2} g h^2\right) + \frac{\partial}{\partial y} (u v h) &=& 0  \label{swe} \\
 \frac{\partial}{\partial t} (vh) + \frac{\partial}{\partial x} (u v h) + \frac{\partial}{\partial y} \left(v^2 h + \frac{1}{2} g h^2\right) &=& 0 \;.
\nonumber
\end{eqnarray}
Here $h(t,x,y)$ is the fluid layer thickness, and $u(t,x,y)$ and $v(t,x,y)$ are the components of 
the velocity field of the fluid. The gravitational acceleration is denoted by $g$. 

We consider a spatial domain $\Omega = [-3,\,3]^2$ (spatial units), 
and an integration window is $t_0 = 0 \le t \le t_\textrm{f} = 0.1$ (time units). 
Boundary conditions are considered periodic. The space discretization is realized using a finite volume scheme, 
and the time integration uses a fourth Runge-Kutta scheme, following the method Lax-Wendroff \cite{Wendroff_1998}.
The model uses a square -$q \times q$ uniform spatial discretization grid, which brings the number of model (state) variables to $n = 3\,q^2$.

We use the automatic differentiation tool TAMC \cite{giering1997tangent,TAMC_1998} 
to build various sensitivity models, as follows. 
The tangent-linear model ({\sc tlm}) propagates perturbations forward in time. 
The first-order adjoint model ({\sc foa}) propagates perturbations backwards in time, 
and efficiently computes the gradient of a scalar cost functional defined on the model states. 
The second-order adjoint model ({\sc soa}) computes the product between the Hessian of the cost function 
and a user-defined vector \cite{Cioaca_2011}. 

The overhead introduced by the sensitivity models is considerable.
Table \ref{Table:CPUTimes_exp} presents the CPU times of {\sc tlm}, {\sc foa}, and {\sc soa} models 
(normalized with respect to that of one {\sc fwd} run). 
One {\sc soa} integration is about $3.5$ times more expensive than a single first-order adjoint run, 
while the {\sc foa} takes $3.7$ times longer than the forward run. 
These relative costs depend on the discretization methodology and the implementation of sensitivities. 
Our previous research showed how to build efficient adjoint models by reusing computations performed
in the forward run \cite{Cioaca_2011}. For example, for the shallow water model, the alternative continuous adjoints we built required
a fraction of the forward model CPU time to run. 

\begin{table}
\caption{Normalized CPU times of different sensitivity models. The forward model takes one time unit to run.}
\centering
\begin{tabular}{ll|ll}
\hline
  {\sc fwd} & $1$ & & \\ \hline
  {\sc tlm} & $2.5$ & {\sc fwd} + {\sc tlm} & $3.5$ \\ \hline
  {\sc foa} & $3.7$ & {\sc fwd} + {\sc foa} & $4.7$\\ \hline
  {\sc soa} & $12.8$ & {\sc fwd} + {\sc tlm} + {\sc foa} + {\sc soa} & $20$ \\ \hline
\end{tabular}
\label{Table:CPUTimes_exp}
\end{table}

\subsection{Data assimilation setup}\label{sec:dasetup}

\begin{figure}
\centering
\subfigure[Initial time]{ \includegraphics[trim=38 20 30 17, clip, width=0.4\textwidth]{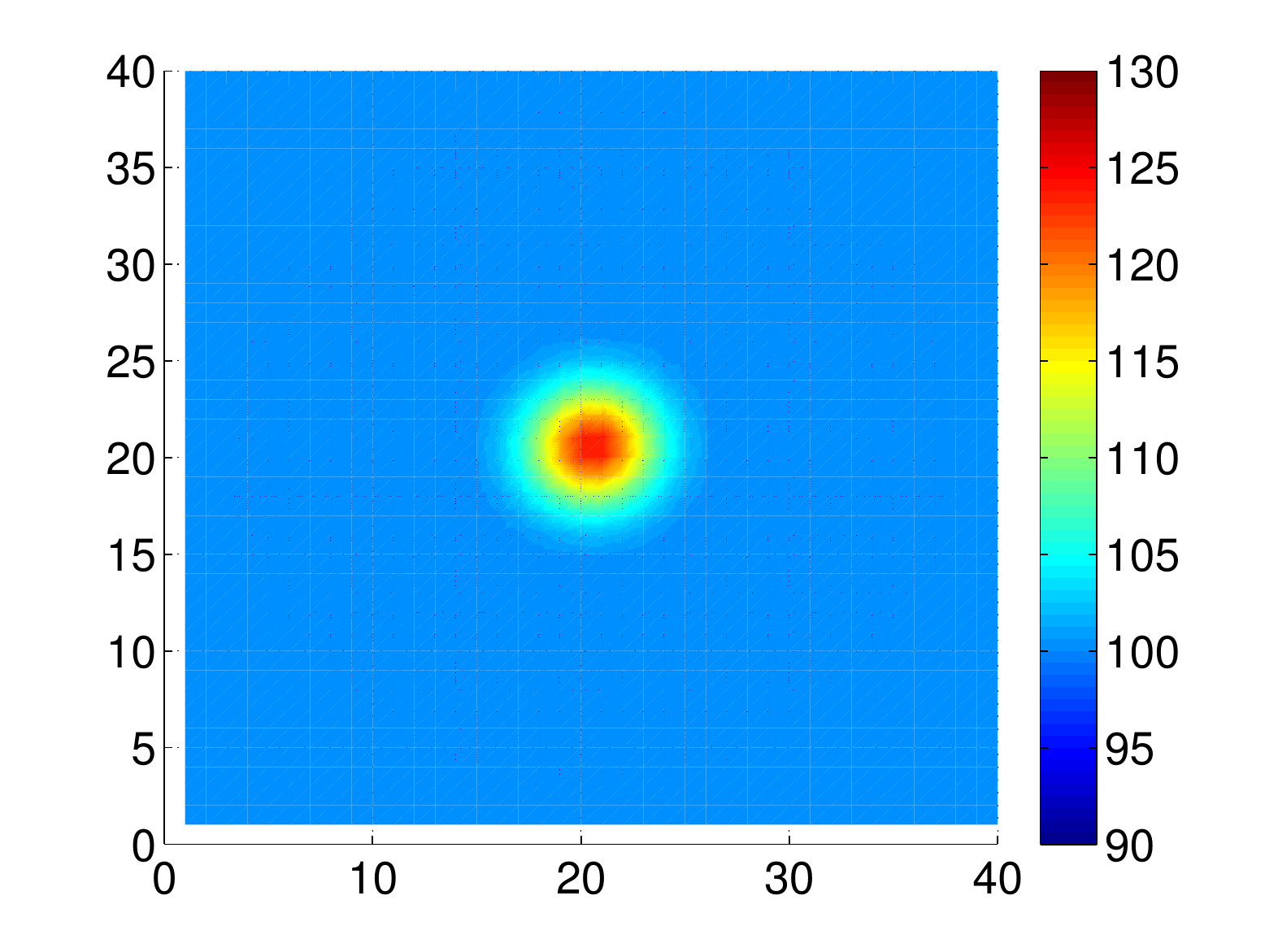} }
\subfigure[Final time]{ \includegraphics[trim=38 20 30 17, clip, width=0.4\textwidth]{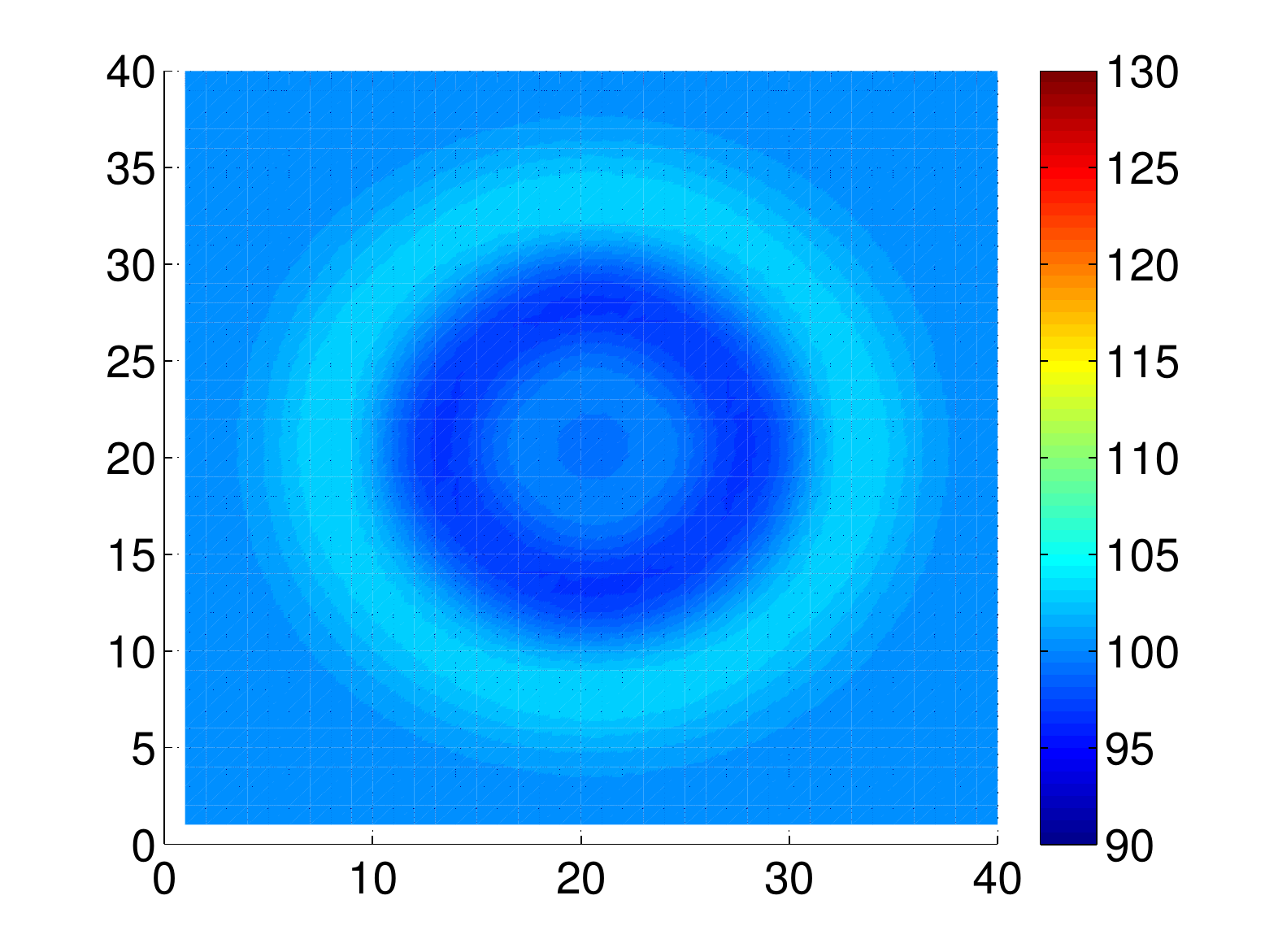} }
\caption{The height field $h$ at the beginning and at the end of the reference trajectory.}
\label{fig:swetraj}
\end{figure}

The 4D-Var system is set up for a simple version of the ``\textit{circular dam}'' problem \cite{anastasiou1997solution}.
The reference initial height field $h$ is a Gaussian bell of a width equal to $1$ length units centered on the grid, and 
the reference initial velocity vector components are constant $u=v=0$. 
The physical interpretation is that the front of water falls to the ground ($h$ decreases) under the effect of gravity
and creates concentric ripples which propagate towards the boundaries. 
Figures \ref{fig:swetraj} represent snapshots of the reference trajectory at initial and final time. 

The computational grid is square and regular with $q=40$ grid points in each direction, for a total of $4800$ model variables (states).
The simulation time interval is set to $0.01$ seconds, using $N=100$ timesteps of size $0.0001$ (time units).

The $h$ component of the a priori estimate (background) $\xb$ is generated by adding a correlated perturbation to the $h$ reference solution at initial time.
The background error covariance $\B_0$ corresponds to a standard deviation of $5\%$ of the reference field values.
For the $u$ and $v$ components we use white noise to prescribe perturbations.
The spatial error correlation uses a Gaussian decay model, with a correlation distance of five grid points. 

Synthetic observations are generated at the final time, by adding random noise to the reference trajectory.
Since the observation errors are assumed uncorrelated, the observation error covariance matrix $\R$ is diagonal. 
The standard deviation for observation noise is $1\%$ of the largest absolute value of the observations for each variable.
The observation operator $\Hobs$ is linear and selects observed variables at specified grid points. 
For the following experiments, we consider observations of all variables at each grid point. 

The minimization of the 4D-Var cost function is performed with the L-BFGS-B solver \cite{zhu1997algorithm} using a fixed number of $100$ iterations.

\subsection{Experimental results}

\subsubsection{Validating the low-rank approximations of the observation impact matrix}

In the first experiment, we study the effect of two possible sources of errors: data noise and SVD truncation errors.
These issues are inherent to performing the data assimilation and the observation impact analysis.

\begin{figure}
\setcounter{subfigure}{0}
\centering
 \subfigure[$h$]{ \includegraphics[width=0.3\textwidth]{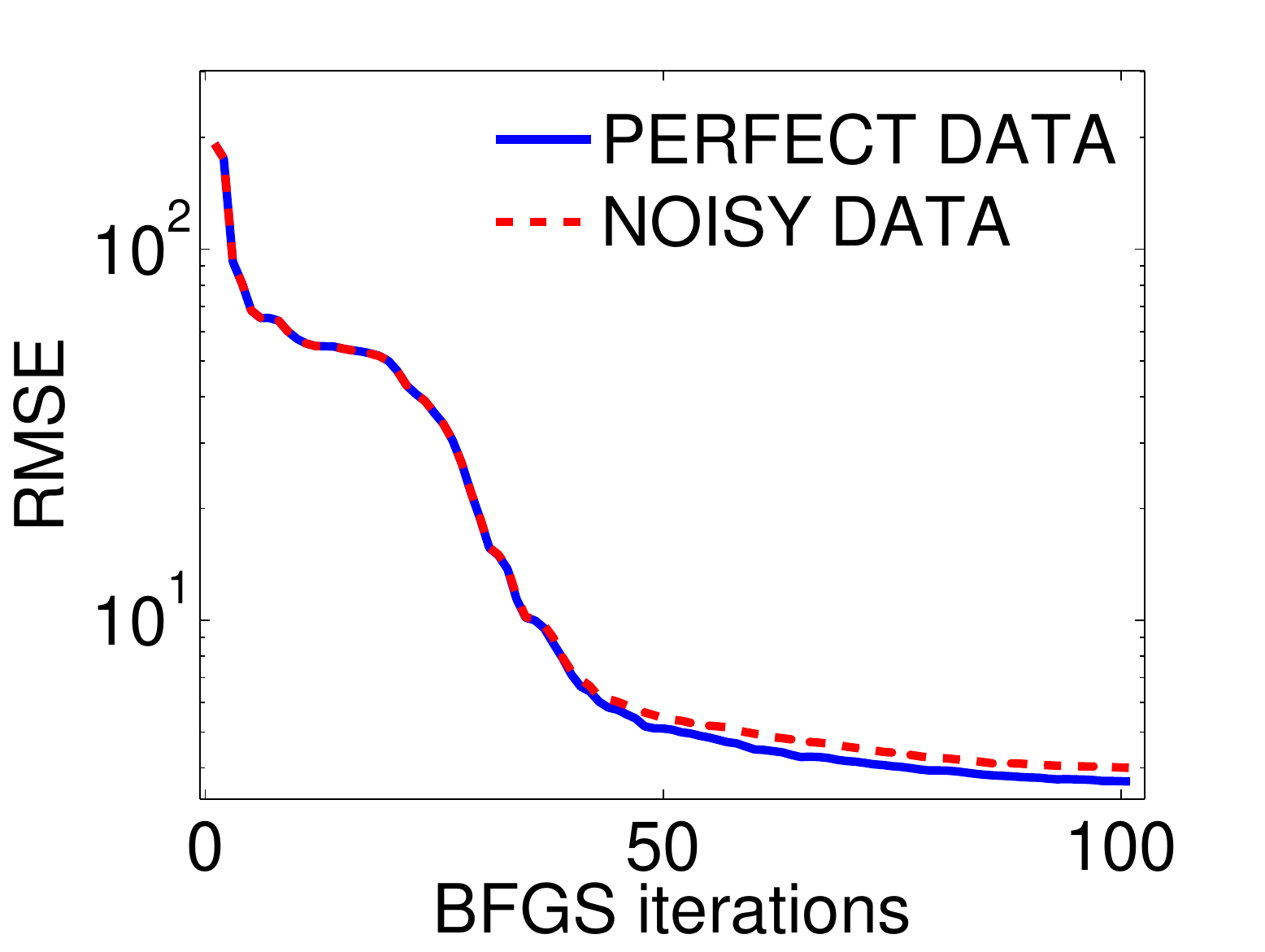} }
 \subfigure[$u$]{ \includegraphics[width=0.3\textwidth]{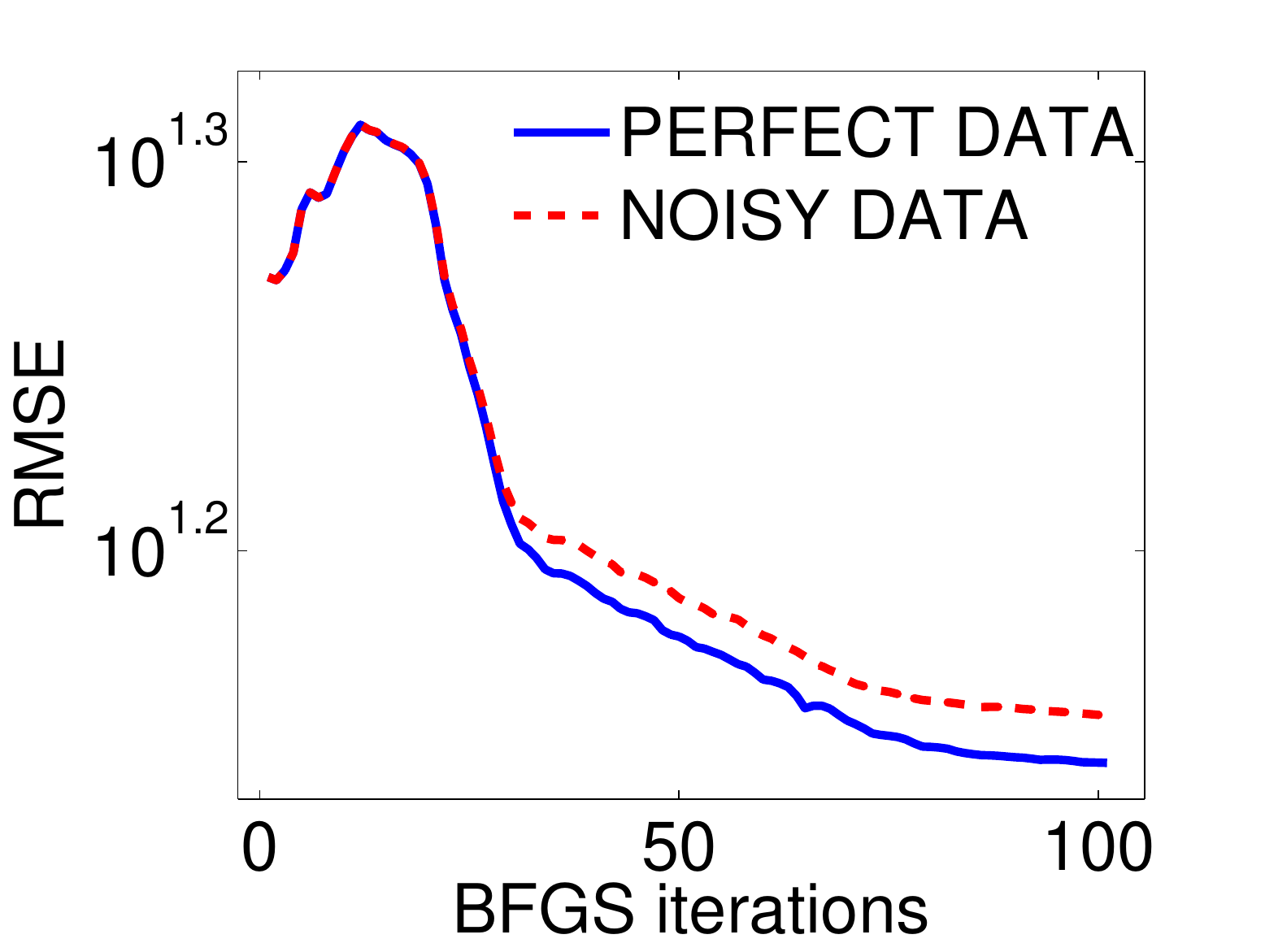} }
 \subfigure[$v$]{ \includegraphics[width=0.3\textwidth]{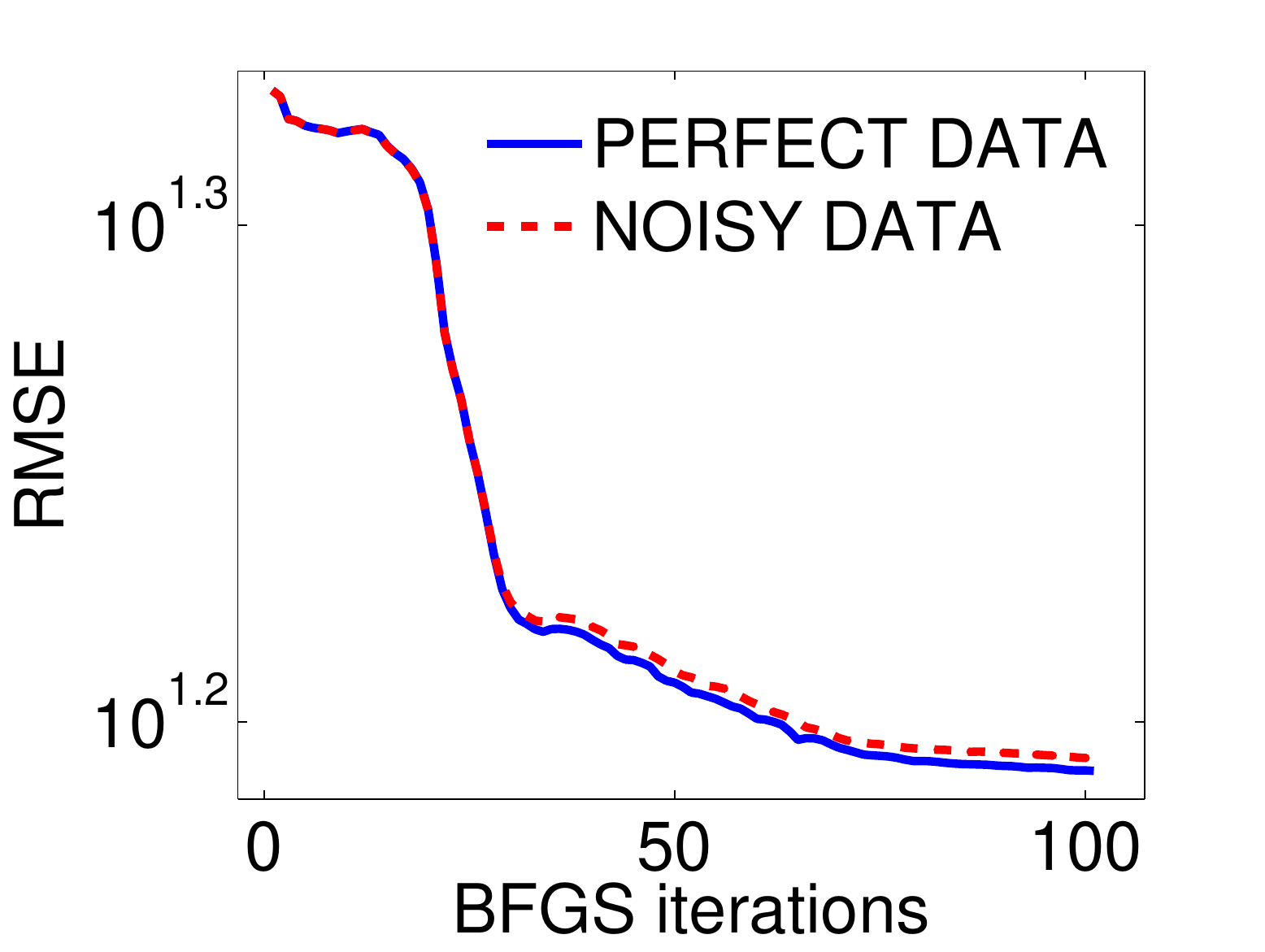} }
\caption{Root mean square error decrease for each variable versus the number of L-BFGS iterations.}
 \label{fig:4dvar_perfvspert}
\end{figure}

We apply our computational methodology to the data assimilation scenario introduced in Section \ref{sec:dasetup}. 
To assess how the reanalysis is affected by the presence of noise,
we first assimilate perfect observations (i.e., reference model values), then assimilate the same observations with added small noise. 
Figure \ref{fig:4dvar_perfvspert} reveals that the convergence of the numerical optimization procedure is similar in the two scenarios,
and leads to a similar decrease in the root mean square (RMS) error for each variable.


We now apply the systematic approach presented in Section \ref{sec:lowrank} to each one of the two assimilation scenarios,
in order to compute the sensitivity to observations \eqref{eqn:ST0_psigrad}  when the verification solution $\xv_0$ is 
the background $\xb_0$ and $\C$ is the identity matrix:
\begin{align}
  \nabla_{\y_k} \Psi(\xa_0) = \T\, (\xa_0 - \xb_0 )\,.
 \label{eqn:OI_invobsimp}
 \end{align}
This procedure provides the sensitivity of the 4D-Var increment norm $\left\Vert \xa_0 - \xb_0 \right\Vert$ to data.
Large sensitivity values will be associated to observations that had a more important contribution to the 4D-Var correction.

\begin{figure}
\setcounter{subfigure}{0}
\centering
 \subfigure[perfect $h$ observations]{ \includegraphics[width=0.4\textwidth]{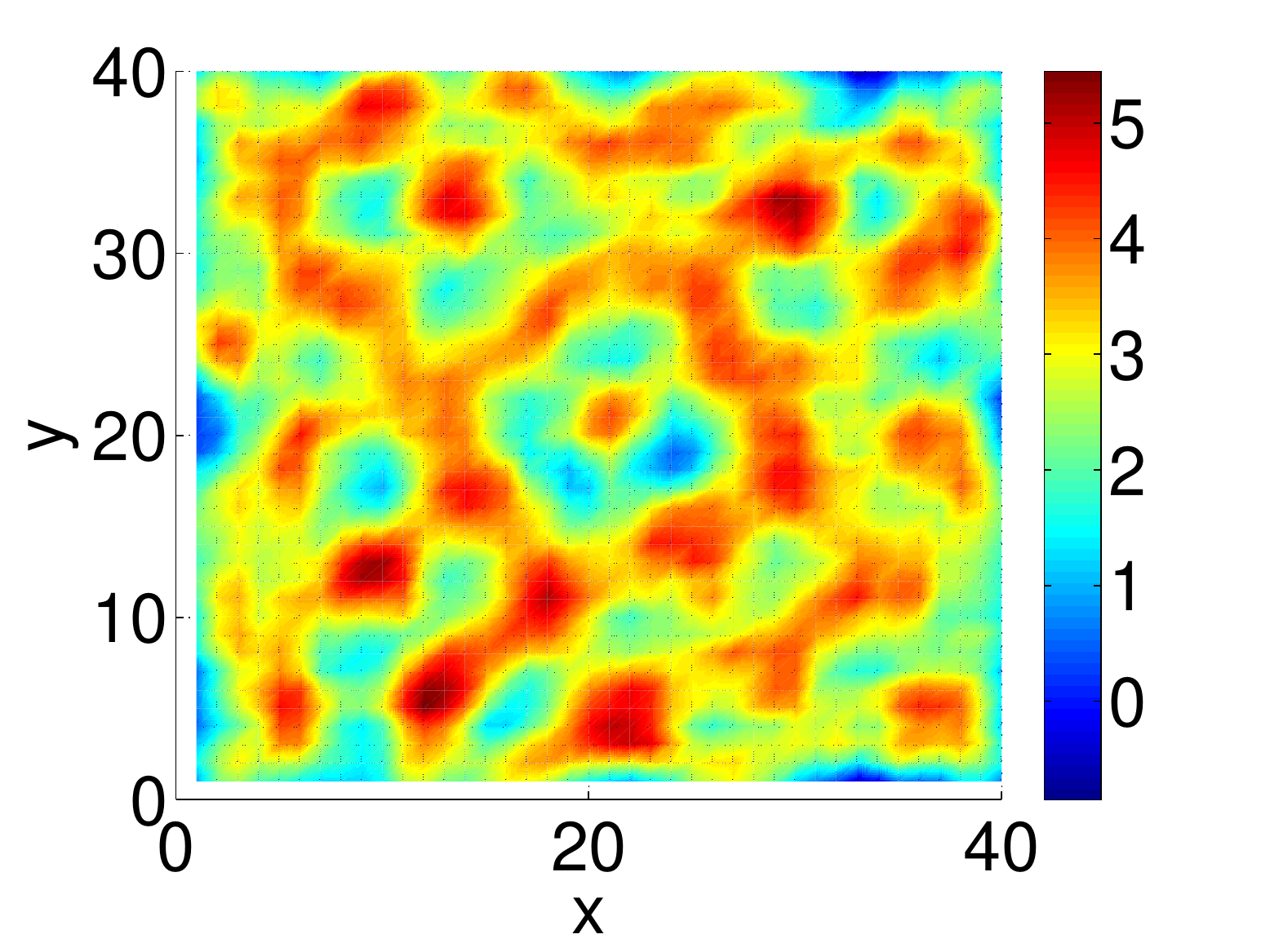} \label{fig:OI_perfvspert_h}}
 \subfigure[noisy $h$ observations]{ \includegraphics[width=0.4\textwidth]{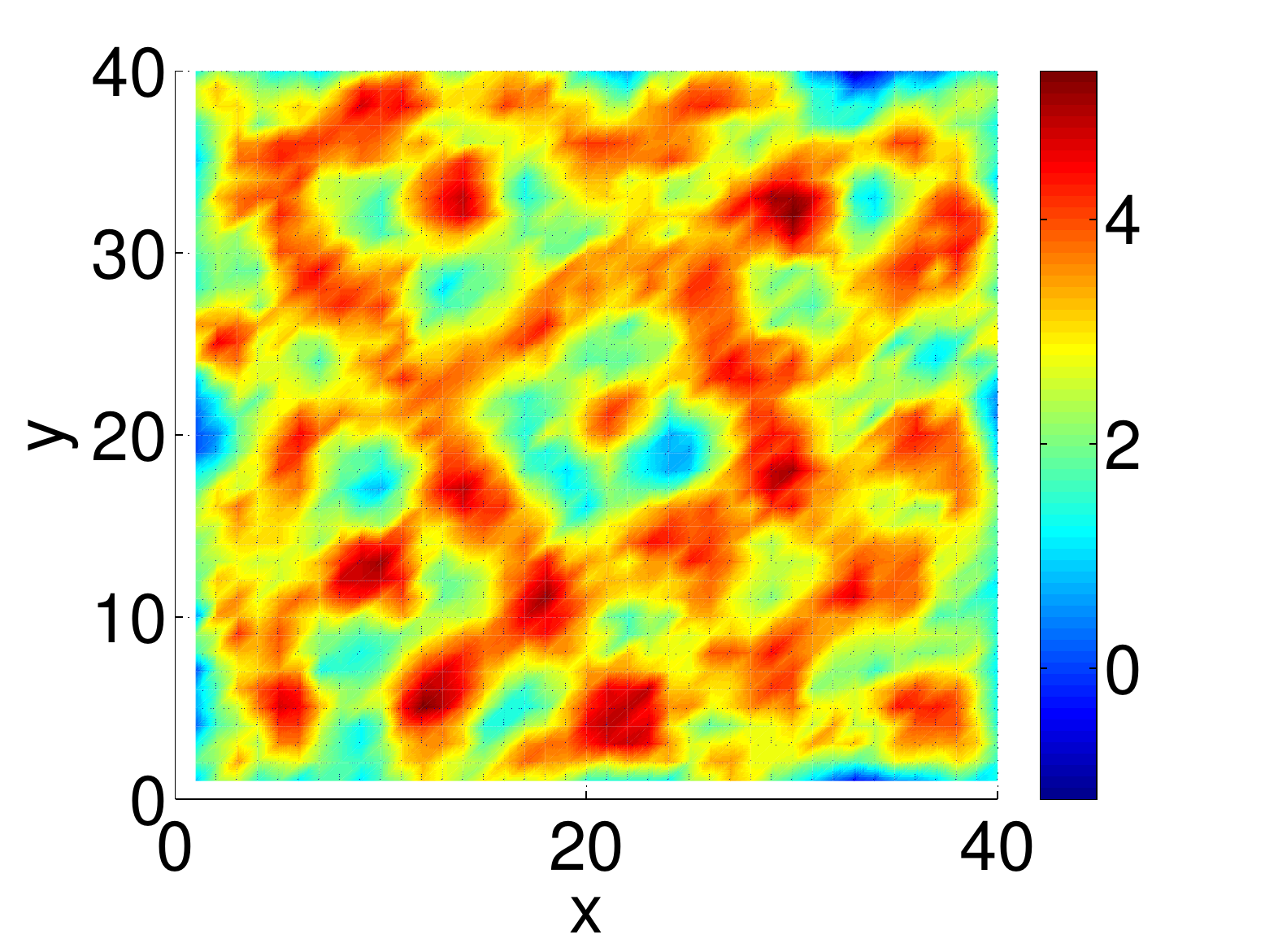} }
 \subfigure[perfect $u$ observations]{ \includegraphics[width=0.4\textwidth]{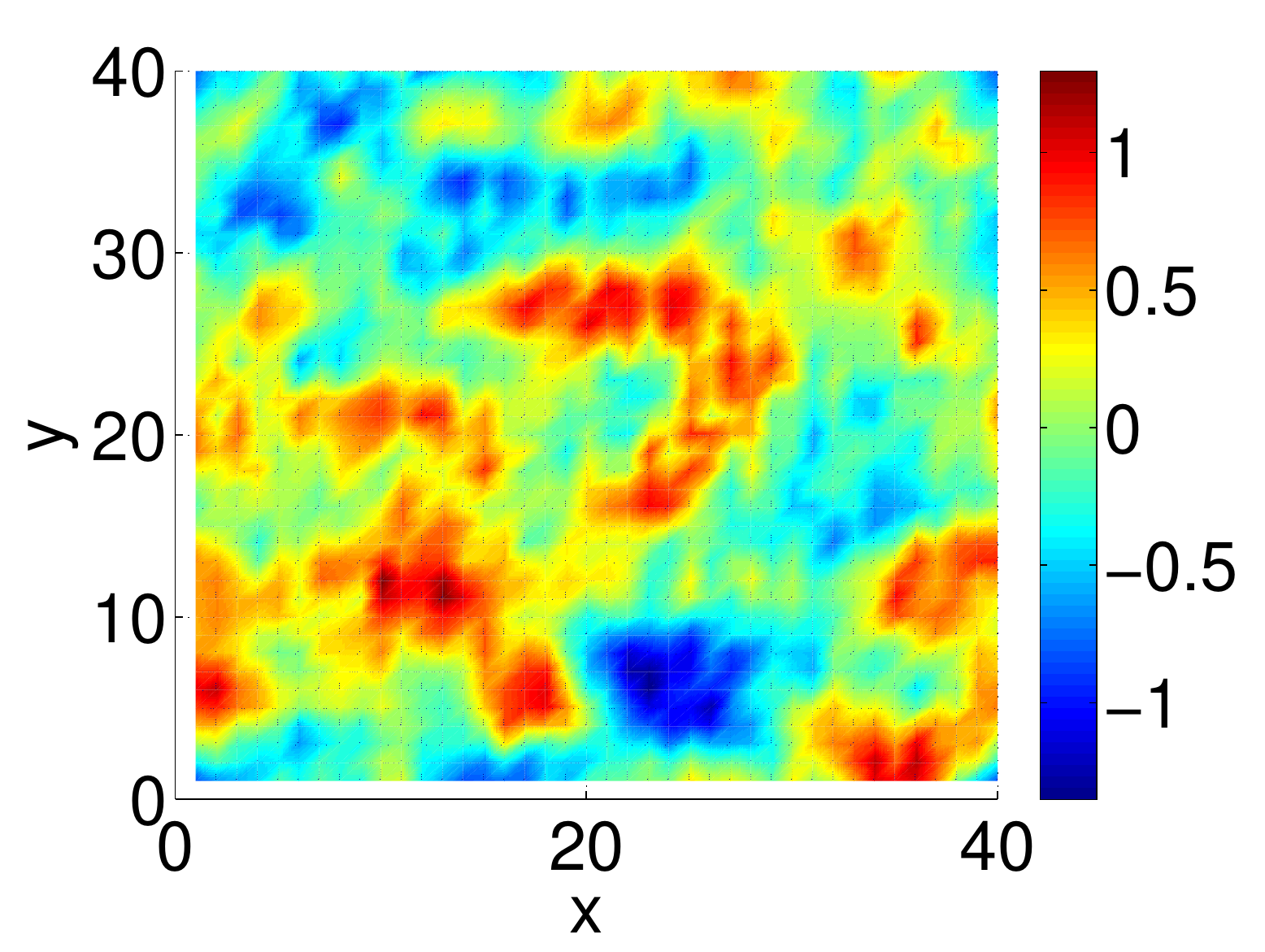} }
 \subfigure[noisy $u$ observations]{ \includegraphics[width=0.4\textwidth]{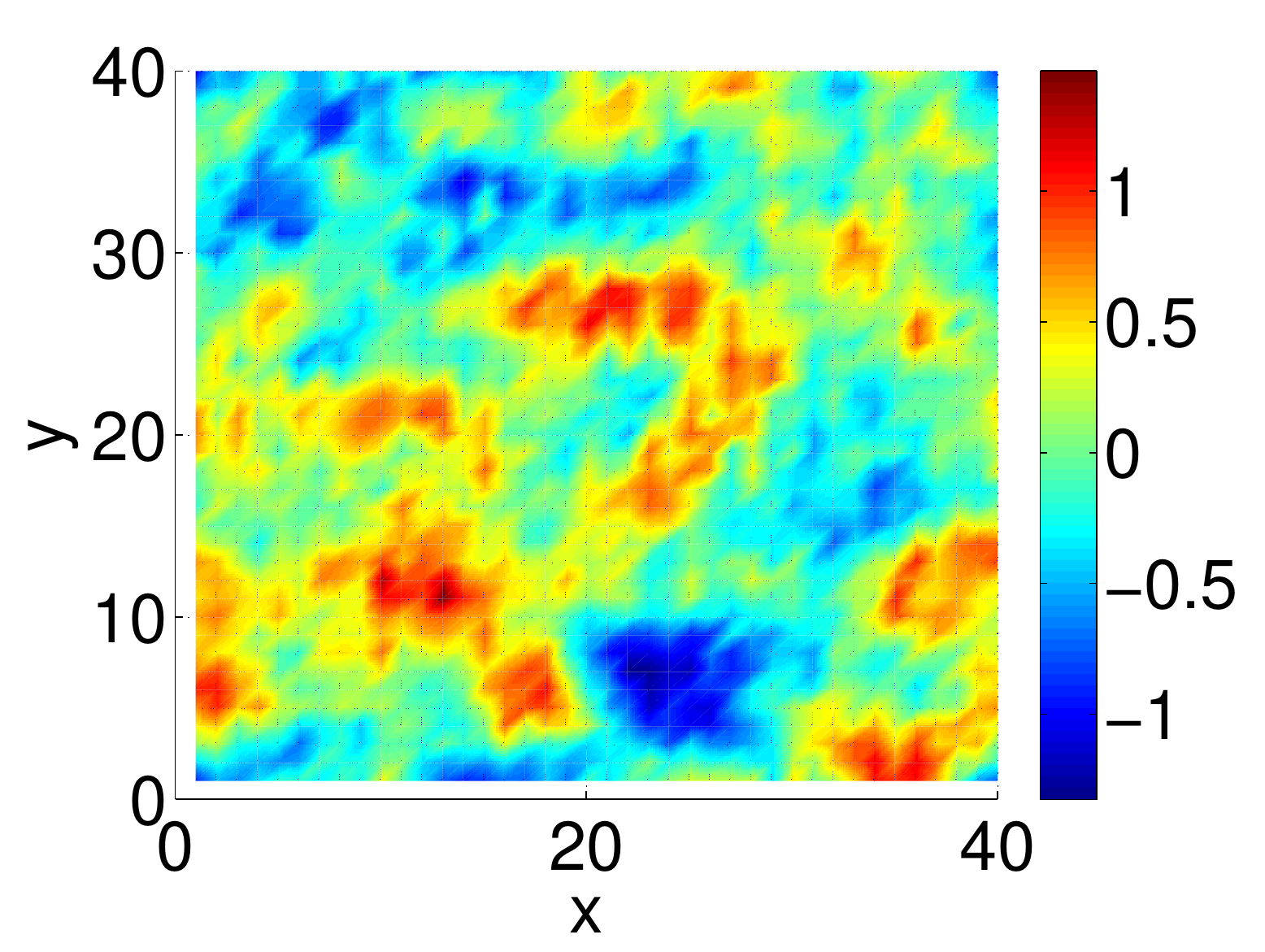} }
 \subfigure[perfect $v$ observations]{ \includegraphics[width=0.4\textwidth]{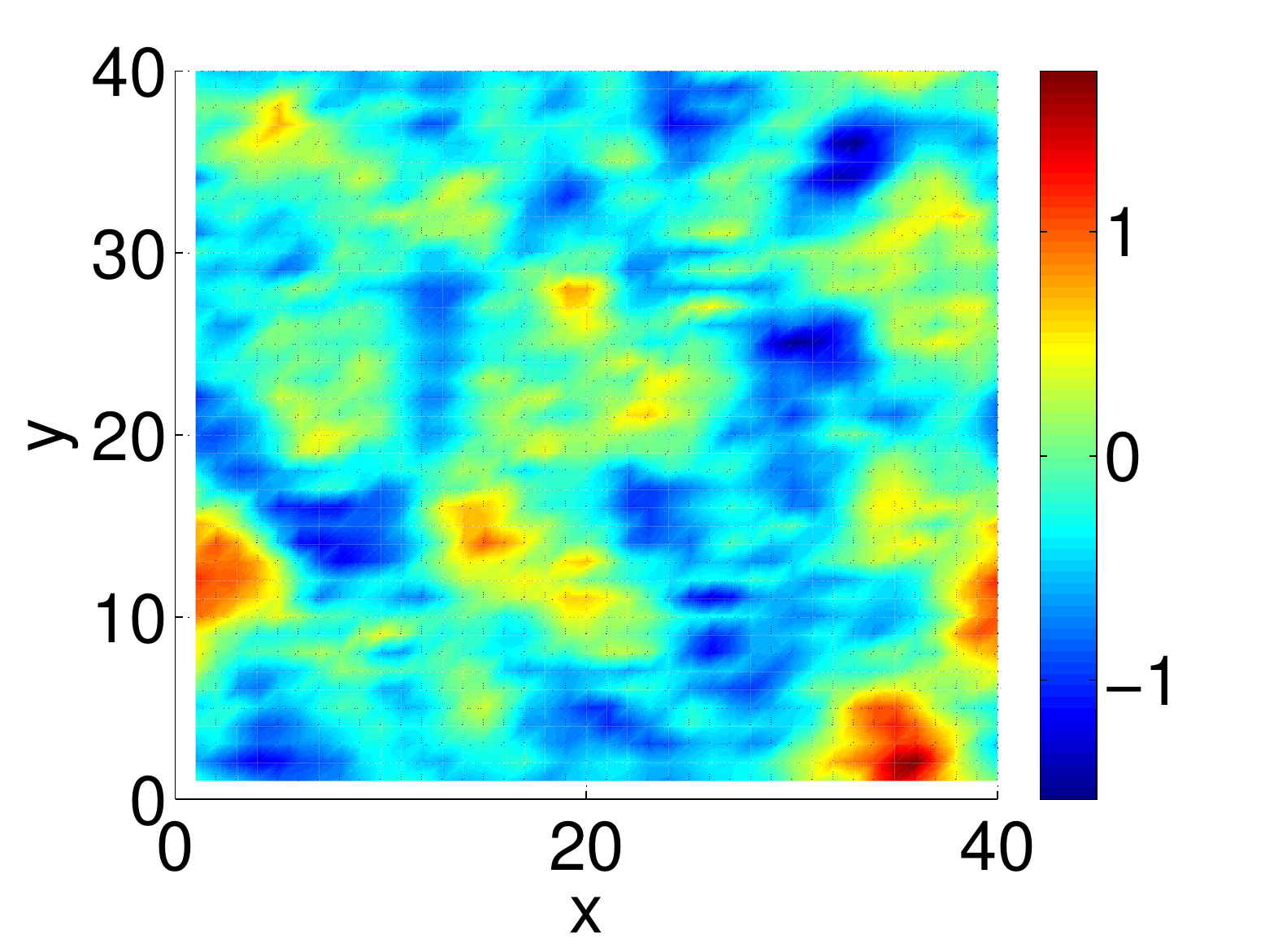} }
 \subfigure[noisy $v$ observations]{ \includegraphics[width=0.4\textwidth]{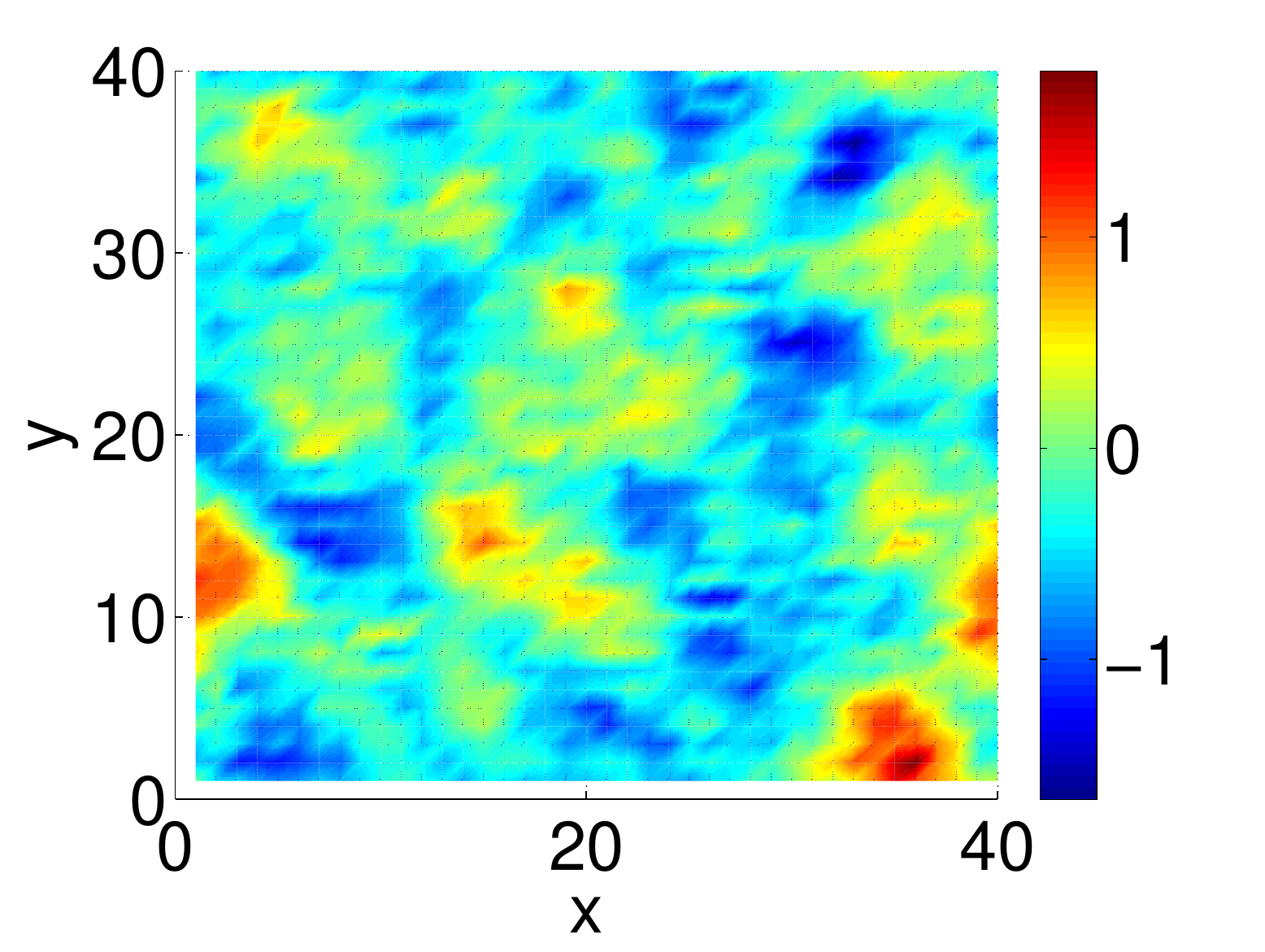} }
  \caption{Sensitivity field \eqref{eqn:OI_invobsimp} for perfect and for noisy observations of $h$, $u$, and $v$.}
 \label{fig:OI_perfvspert}
\end{figure}

Figure \ref{fig:OI_perfvspert} plots the observation sensitivity \eqref{eqn:OI_invobsimp} for each one of $h$, $u$ and $v$, for both perfect and noisy scenarios.
The standard deviation for observation noise is equal to $1\%$ of the largest absolute value of the observations for each variable. 
We notice the observation sensitivity is locally correlated. 
For the $h$ observations $\nabla_{\y_k} \Psi(\xa_0)$ is almost symmetric, while for $u$ and $v$ is aligned along the East-West and South-North directions. 
Observation sensitivities for perfect and for noisy data exhibit similar spatial features and numerical values.
Reasonably small noise in data does not significantly affect the sensitivity fields.

\begin{figure}
 \setcounter{subfigure}{0}
  \centering
  \subfigure[Singular value spectrum]{ \includegraphics[width=0.4\textwidth]{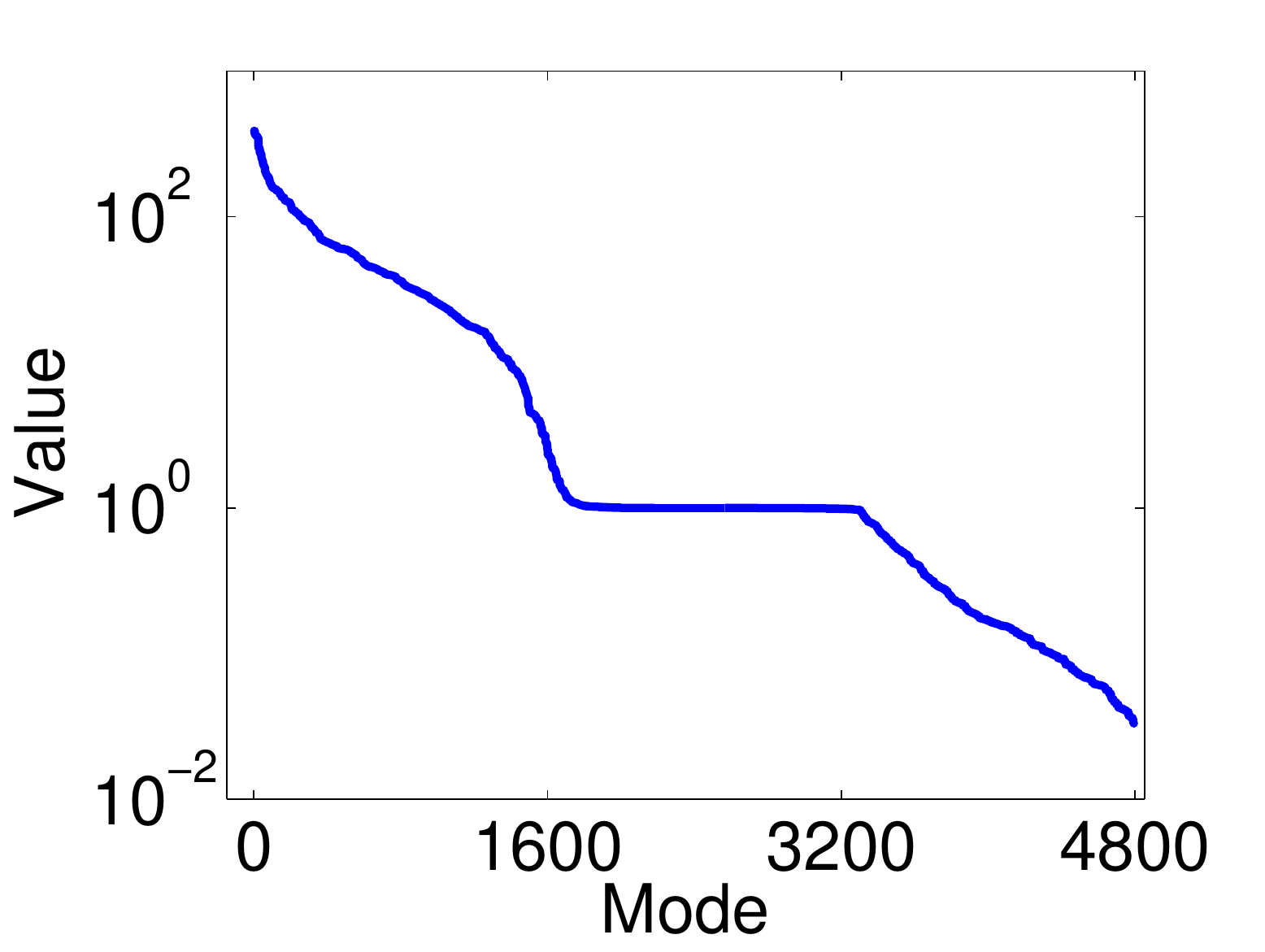} }
  \subfigure[Truncation error]{ \includegraphics[width=0.4\textwidth]{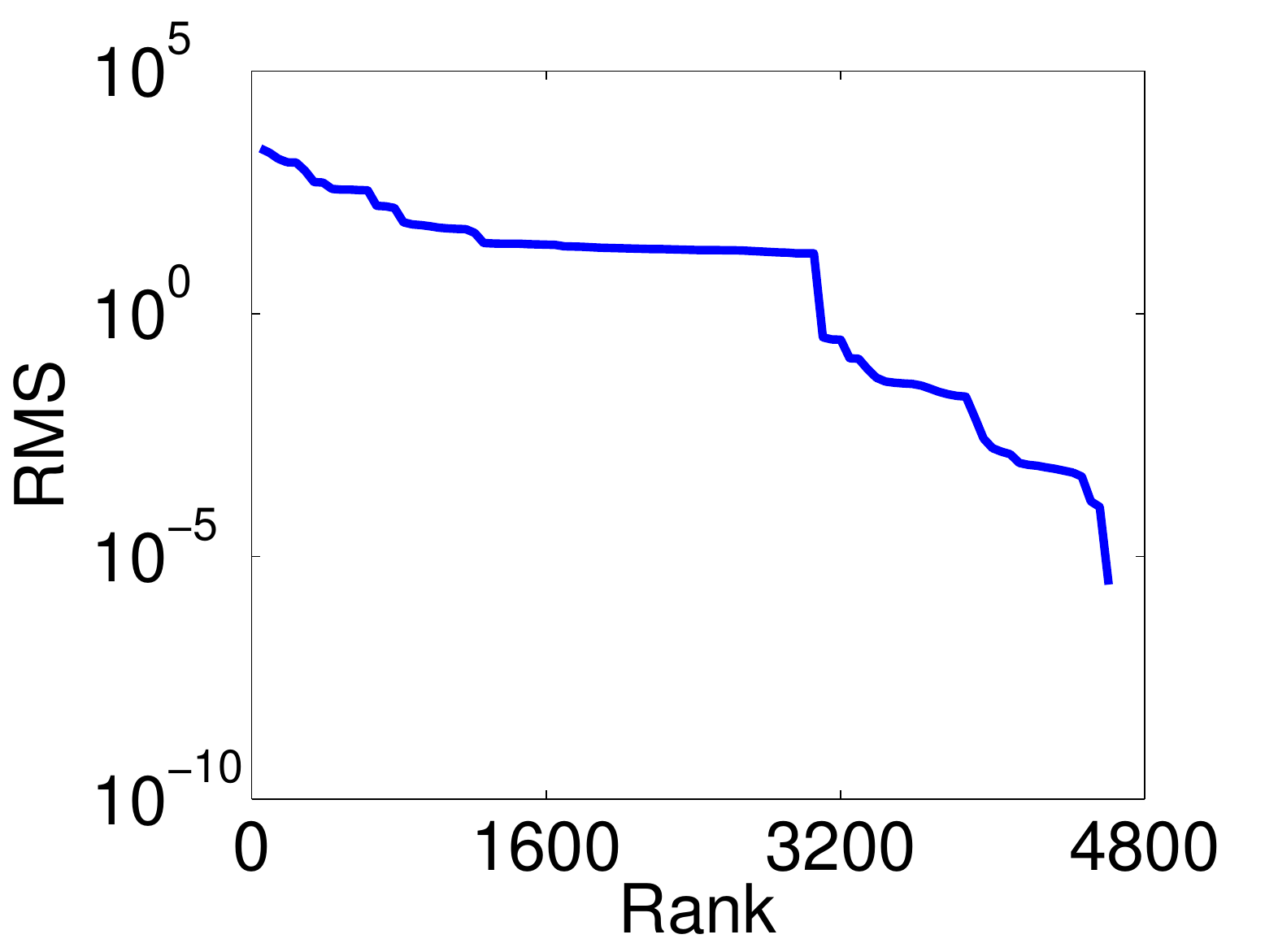} }
  \caption{Singular value decay for the observation impact matrix $\T$ \eqref{eqn:oimatext} and the corresponding truncation error norm.}
  \label{fig:lowrank}
 \end{figure}

The results in Figure \ref{fig:OI_perfvspert} are obtained with the full sensitivity matrix, built offline.
We compare these results with a low-rank approximation of the sensitivity matrix 
computed using Algorithm \ref{alg:iterative}. The low rank approximation of the sensitivity is
\begin{align}
 \left( \nabla_{\y_k} \Psi(\xa_0) \right)_{(p)} = \T_{(p)} \cdot \left( \xa_0 - \xb_0 \right)\,.
 \label{eqn:OI_invobsimplowrank}
\end{align}
Figure \ref{fig:lowrank} displays the singular value spectrum of the observation impact matrix
and the reconstruction error of the low-rank truncated values for observation sensitivity.
It can be noticed that the decay in singular value spectrum resembles that of the truncation error.
Based on this, a good choice for the rank of the approximation is $p \approx 1600$,
which corresponds to one third of the full workload.

\begin{figure}
 \setcounter{subfigure}{0}
  \centering
  \subfigure[Low-rank estimate]{ \includegraphics[width=0.4\textwidth]{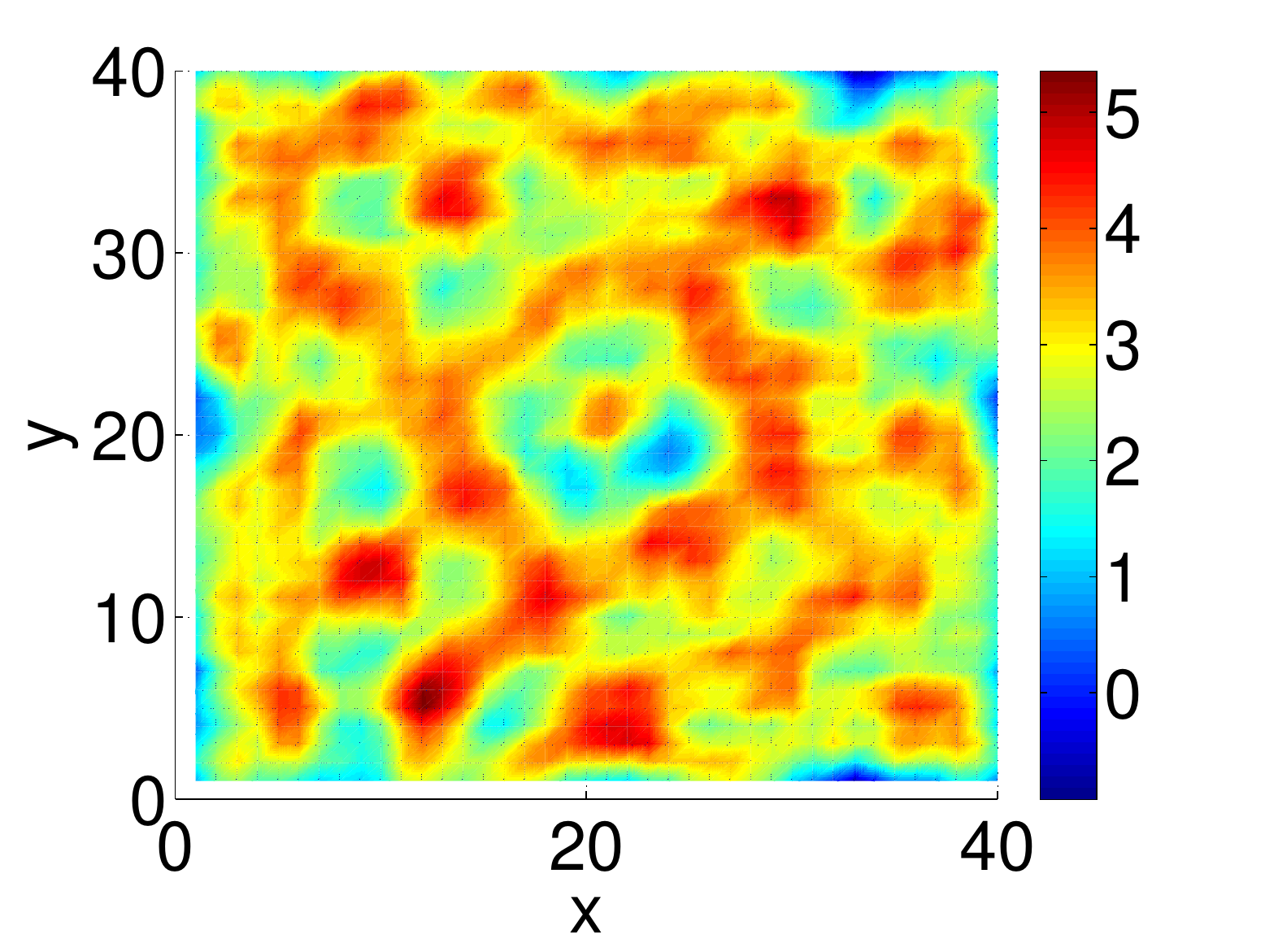} }
  \subfigure[Truncation error field]{ \includegraphics[width=0.4\textwidth]{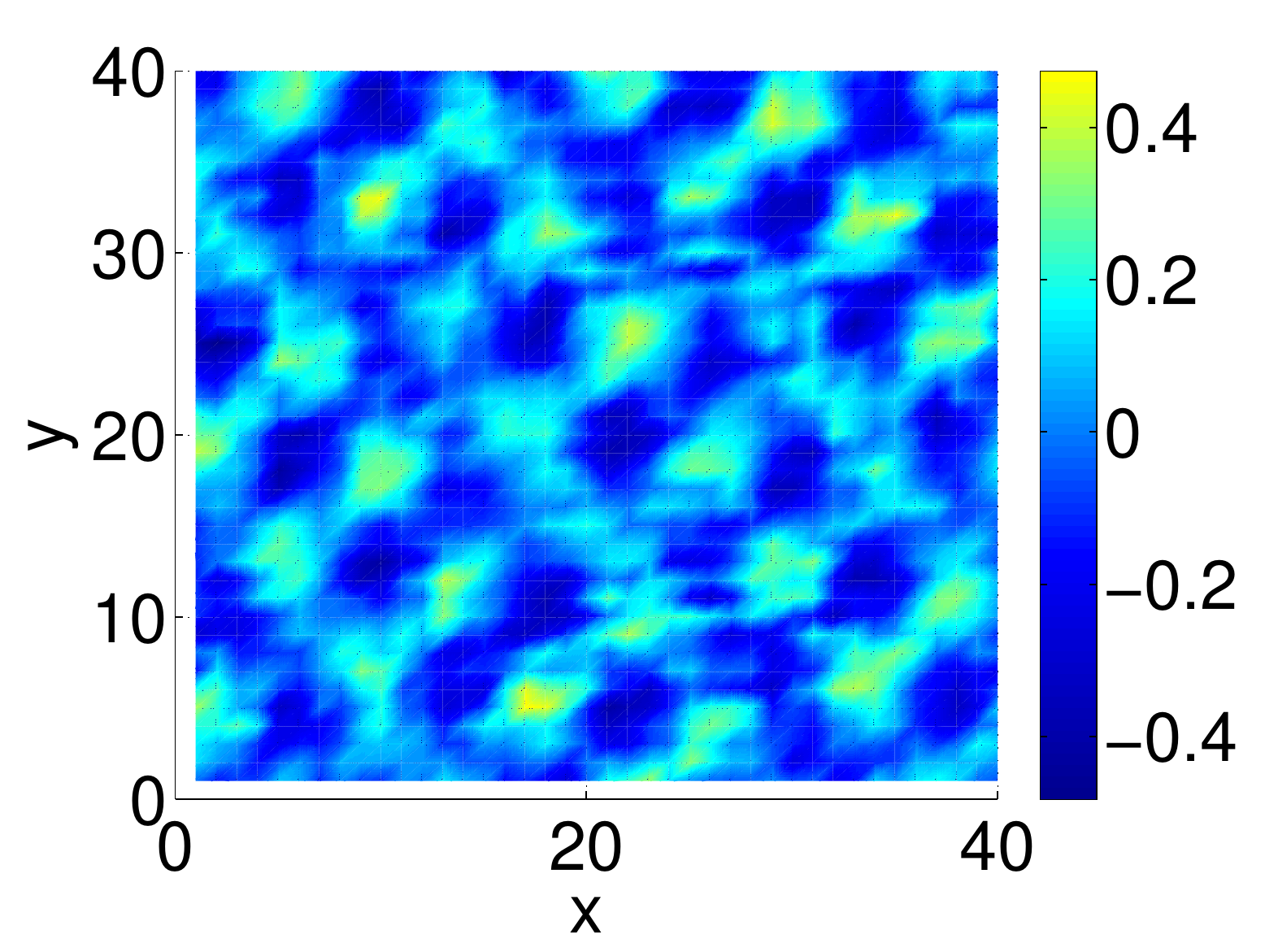} }
  \caption{Low-rank approximation of observation sensitivity \eqref{eqn:OI_invobsimplowrank} for $h$ data and the associated truncation error field for $1600$ modes.}
  \label{fig:OI_redrank}
 \end{figure}

In Figure \ref{fig:OI_redrank} we plot the low-rank approximation \eqref{eqn:OI_invobsimplowrank} of sensitivity to $h$ observations and the corresponding truncation error.
A visual comparison with the full-rank observation sensitivity plotted in Figure \ref{fig:OI_perfvspert_h} reveals that 
the low-rank approximation captures well the main features of the sensitivity field.

\subsubsection{Observation impact}

The computational methodology developed in Section \ref{sec:lowrank} is now applied
to compute the forward observation impact. Specifically, we map changes
in the observations (innovation vector $\Delta \y = \y - \H(\x)$) to changes
in the analysis ($\Delta \xa_0$) using the relation \eqref{eqn:T-mapping}.

\begin{figure}
 \setcounter{subfigure}{0}
  \centering
  \subfigure[Full-rank impact for center observation]{ \includegraphics[width=0.4\textwidth]{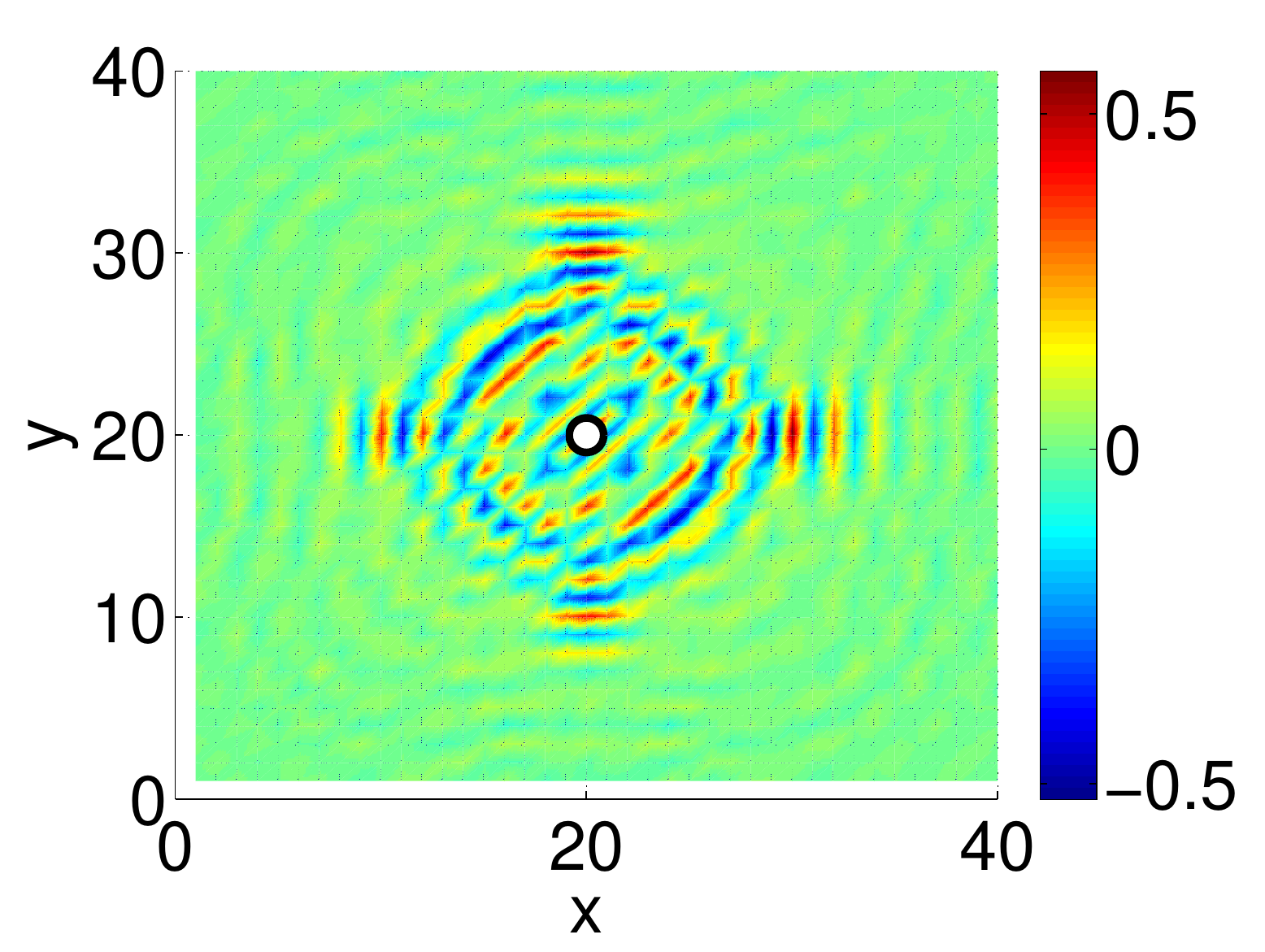} }
  \subfigure[Low-rank impact for center observation]{ \includegraphics[width=0.4\textwidth]{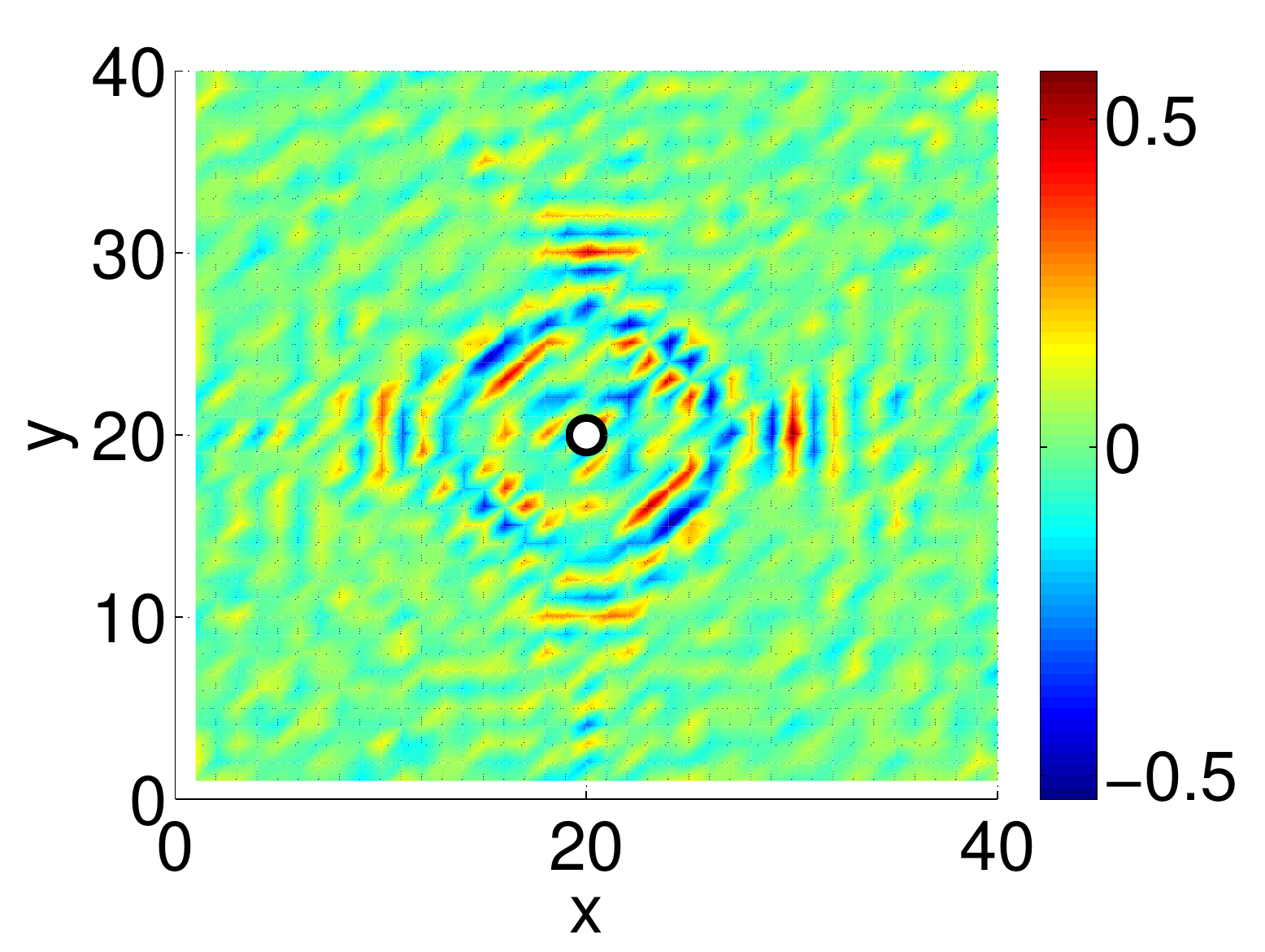} }
  \subfigure[Full-rank impact for corner observation]{ \includegraphics[width=0.4\textwidth]{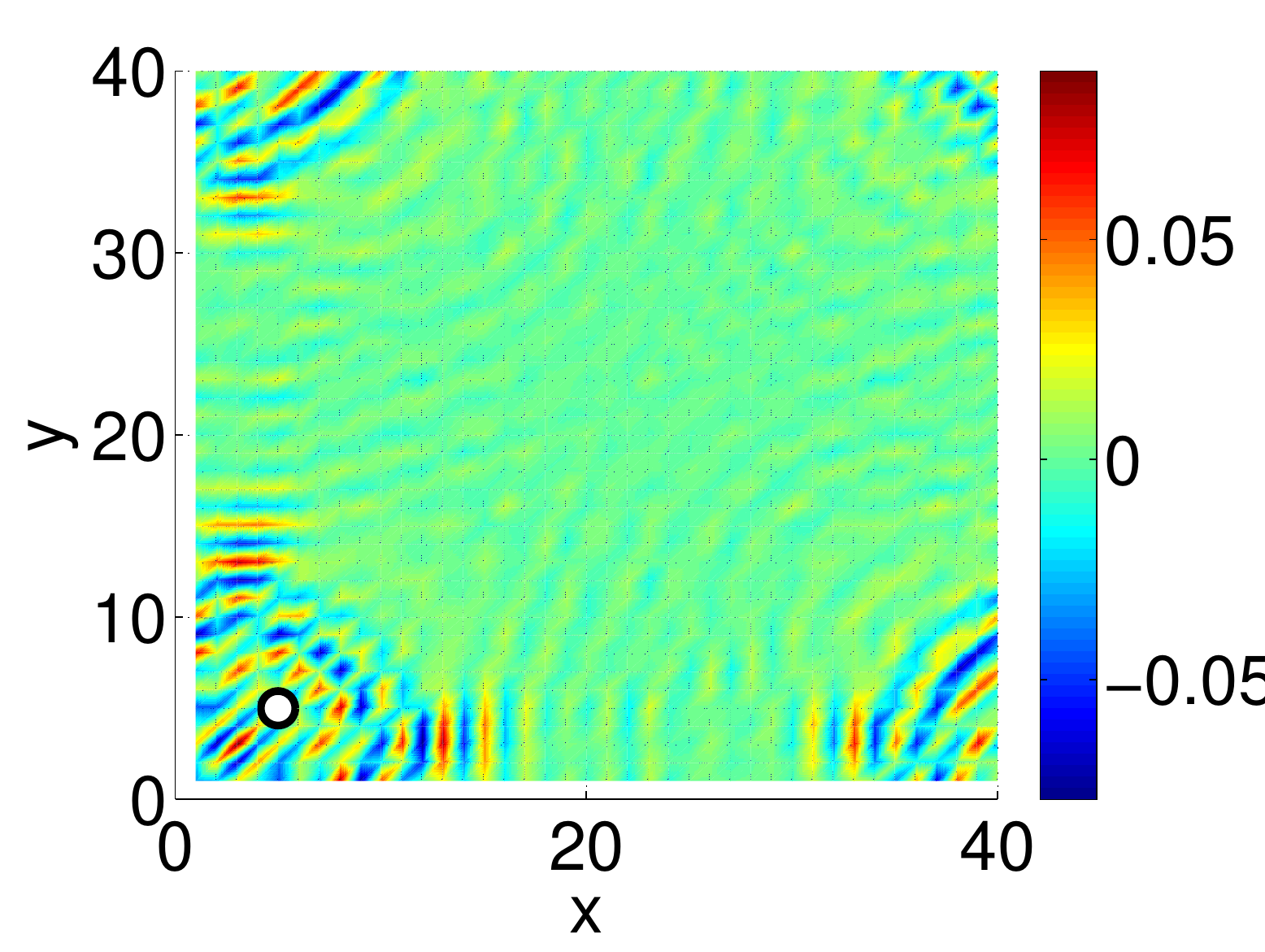} }
  \subfigure[Low-rank impact for corner observation]{ \includegraphics[width=0.4\textwidth]{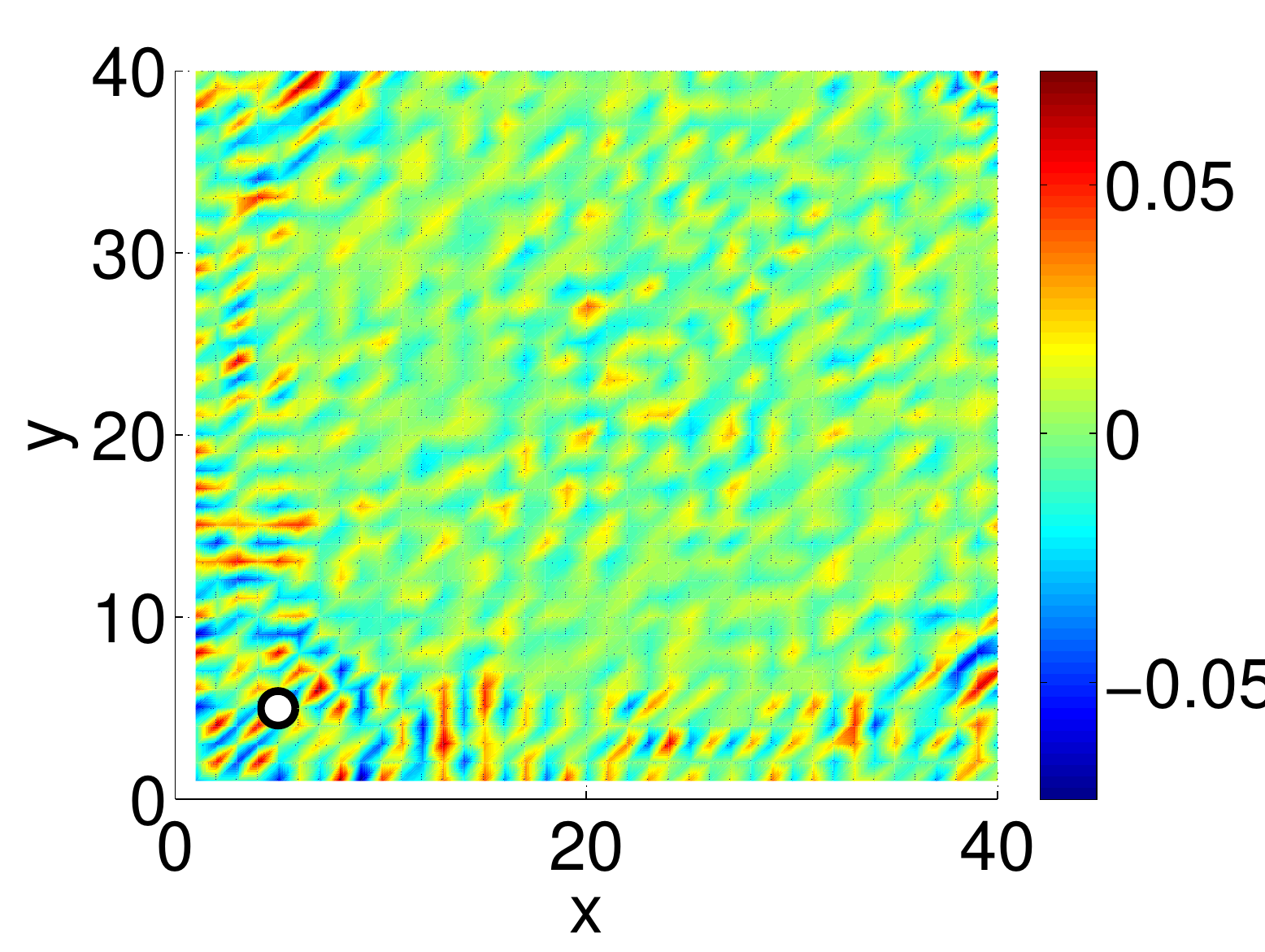} }
  \caption{The impact of two $h$ observations from assimilation time $t_100$, placed in the center and in the corner of the grid.
  Both full rank and reduced rank solutions are shown. $500$ modes are used for the reduced rank approximation.}
  \label{fig:OI_forwardobsimp}
 \end{figure}

We compute the observation impact in full-rank and low-rank corresponding to two observations of $h$,
one in the center (grid coordinates $(20,20)$) and one in the corner (at $(5,5)$) at the locations represented with white markers.
This is achieved  by multiplying $\T^T$ with a vector containing just the innovation brought 
by the observation whose impact we want to evaluate (all other vector entries have value zero).
The resulting impact fields are plot in Figure \ref{fig:OI_forwardobsimp}.
The spatial features of the observation impact have a local radial correlation in both cases.
This means the information carried by the observation is spread in its proximity by the 4D-Var process;
this is also true across the periodic boundaries of the shallow water system.
Moreover, the low-rank approximations are able to pick up the important features of the full-rank calculations,
and provide impacts of a similar magnitude.

\begin{figure}
 \setcounter{subfigure}{0}
  \centering
  \subfigure[First dominant left singular vector (model space)]{ \includegraphics[width=0.4\textwidth]{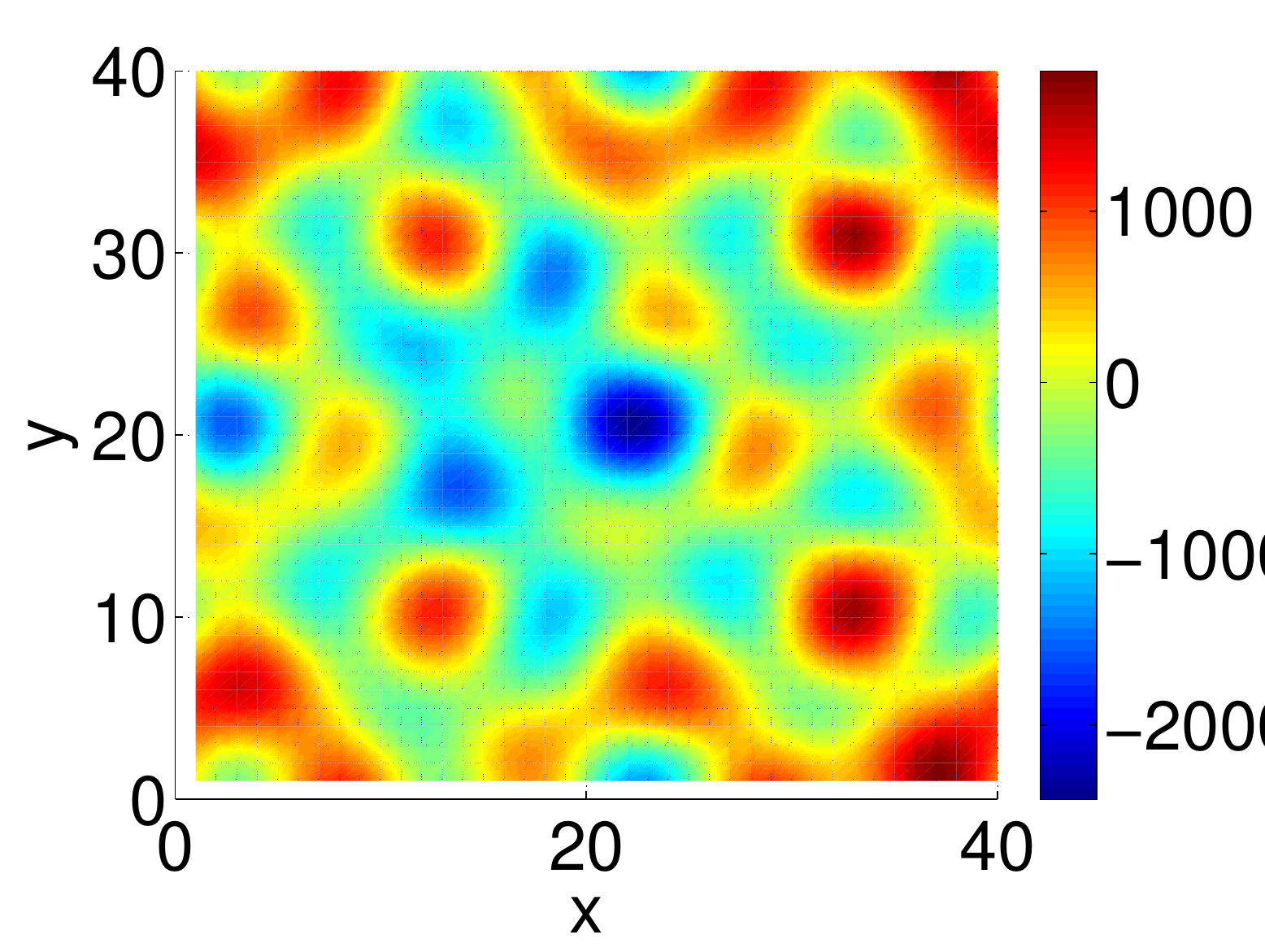} }
  \subfigure[First $500$ dominant left singular vectors (model space)]{ \includegraphics[width=0.4\textwidth]{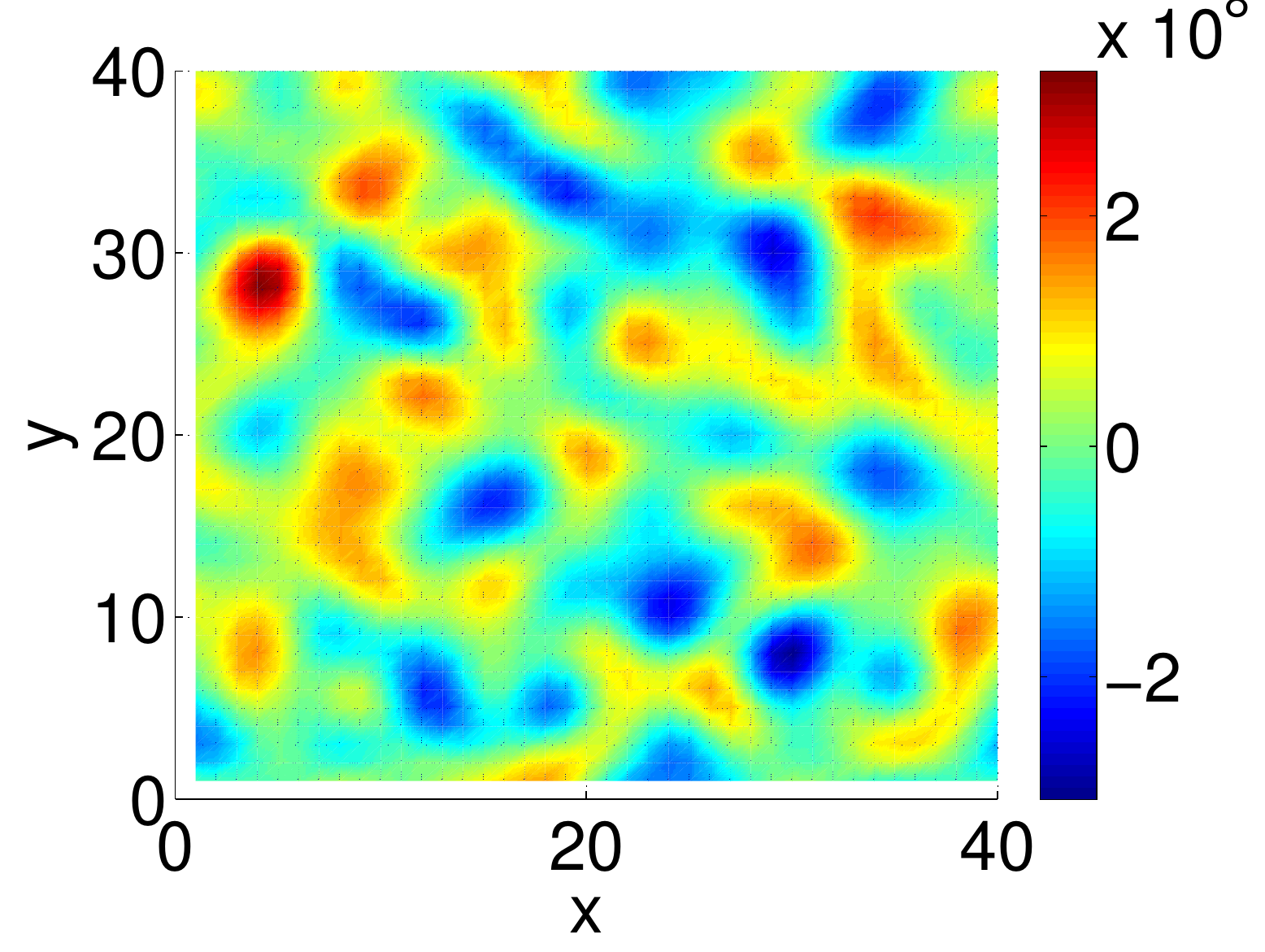} }
  \subfigure[First dominant right singular vector (observation space)]{ \includegraphics[width=0.4\textwidth]{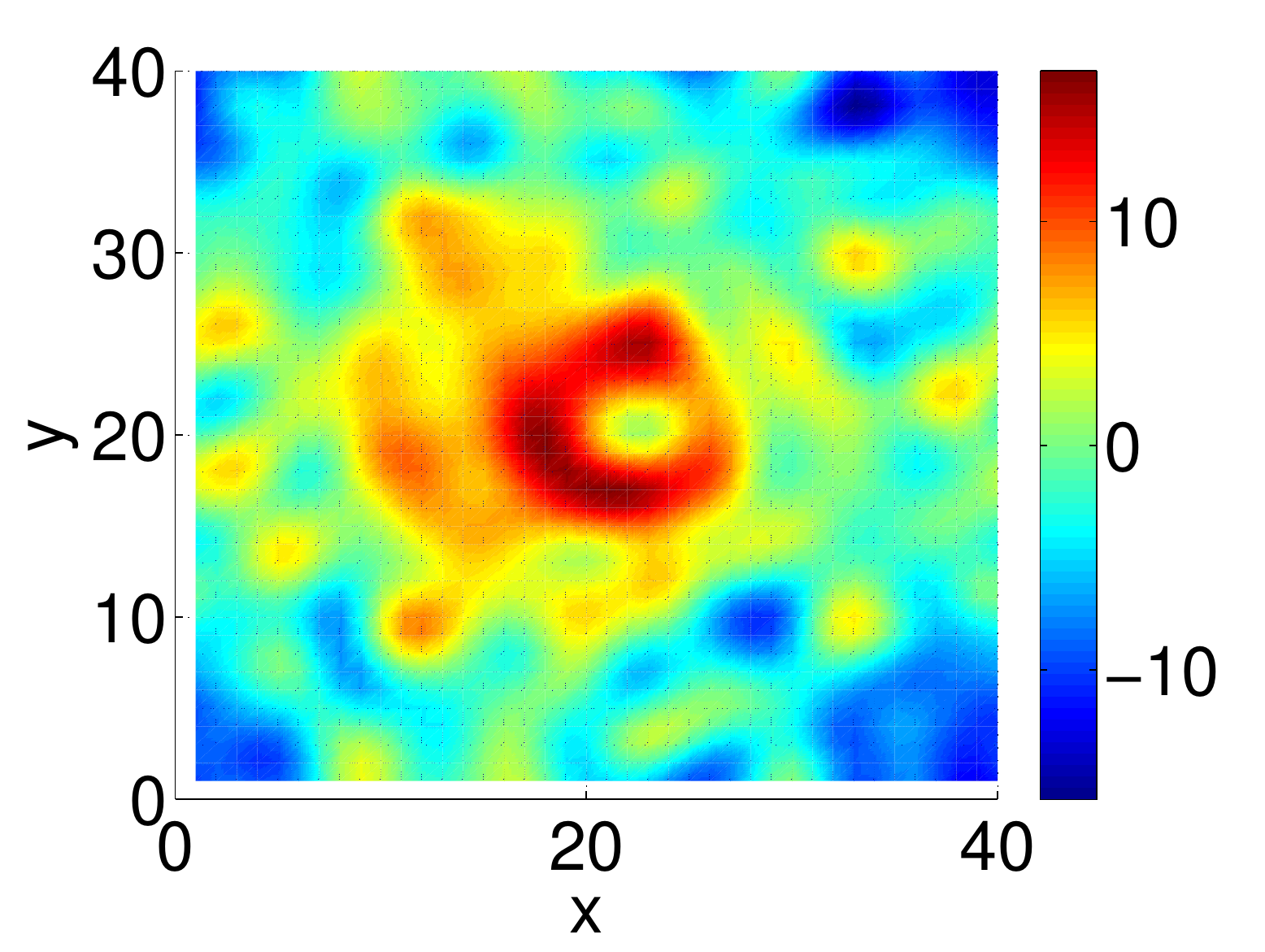} }
  \subfigure[First $500$ dominant right singular vectors (observation space)]{ \includegraphics[width=0.4\textwidth]{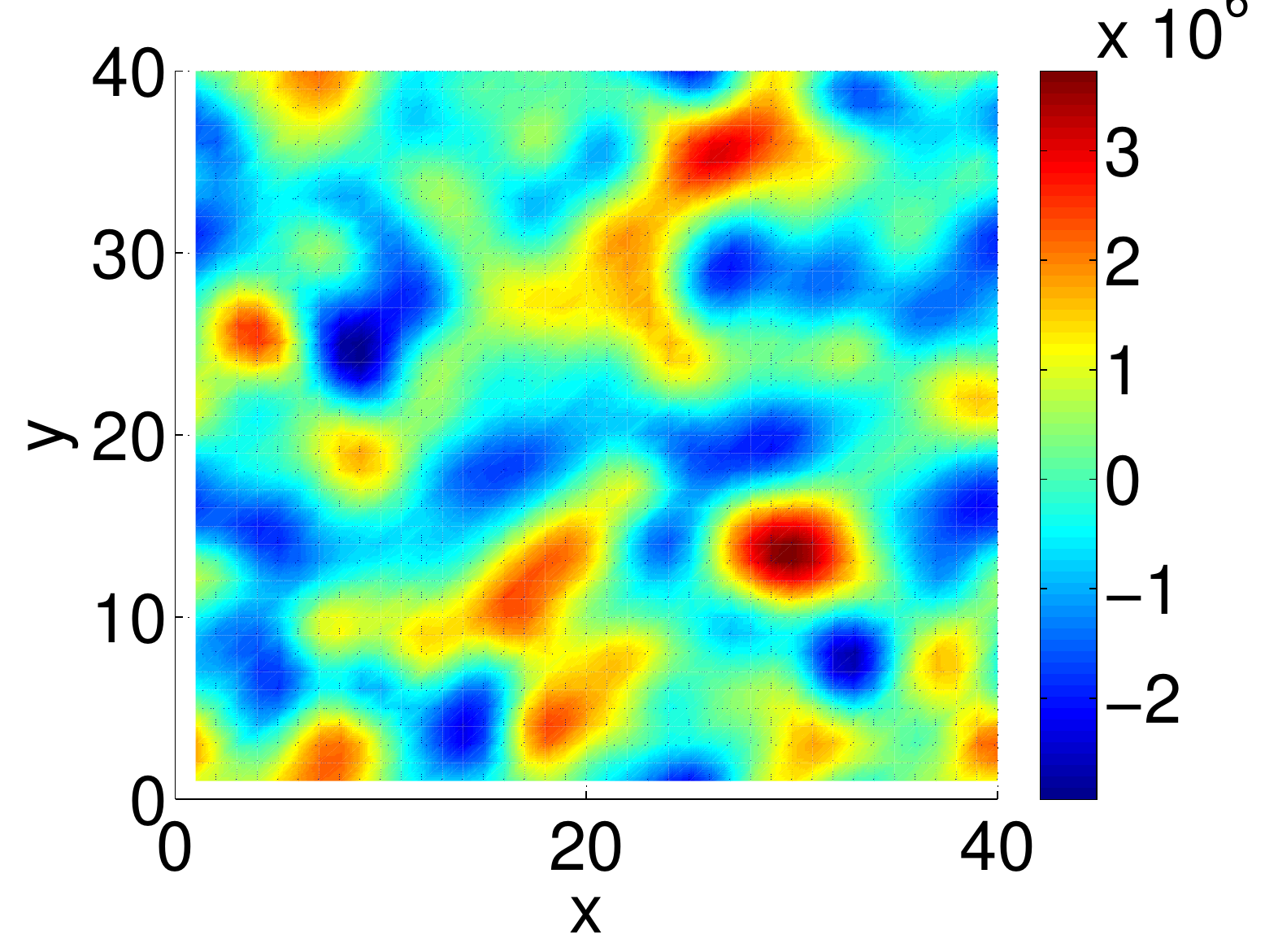} }
  \caption{Principal directions of growth in observation and solution space, as defined by the 
  dominant left and right singular vectors of the observation impact matrix.}
  \label{fig:OI_principaldir}
 \end{figure}

Having already computed the SVD of the observation impact matrix, we also look
at the directions in the data space $\Delta \y$ that have the largest impact on the analysis, and the directions in the 4D-Var correction space $\Delta \x$
that benefit most from the assimilation. These directions are
given by the dominant left and right singular vectors of $\T^T$, respectively.
Figure \ref{fig:OI_principaldir} plots the first dominant left (and right) singular vectors
corresponding to the $h$ variable, and also the composition of the most important $500$ 
directions for the observation and solution space using the formula
\begin{align}
 v_\textrm{domin} = \sum\limits_{i=1}^{500} s_i^2 * v_i\,,
\end{align}

\noindent where $s_i$ and $v_i$ are the singular pair corresponding to the $i$-th mode.

\subsubsection{Pruning observations based on sensitivity values}

In this experiment we illustrate how sensitivity analysis can be used to select the most useful data points.

For this, we compute the 4D-Var reanalysis and then apply the observation impact methodology to compute 
the sensitivity of the cost function $\Psi$ to each one of the $4800$ observations \eqref{eqn:OI_invobsimp}.
We then split our observation set in three subsets of $1600$ observations, corresponding to each one of $h$, $u$ and $v$. 
Within each subset, we order the observations by their sensitivity; the top $800$ observations 
form the {\sc high} sensitivity set, and the ones in the bottom $800$ form the {\sc low} sensitivity set.
This procedure partitions our initial $4800$ observations in two halves with respect to observation sensitivity, taken variable-wise.
The 4D-Var data assimilation process is repeated using either one of the {\sc high} or {\sc low} data sets.

\begin{figure}
\begin{center}
\setcounter{subfigure}{0}
\subfigure[$h$ RMS error decrease versus the number of L-BFGS iterations.]{ \includegraphics[width=0.4\textwidth]{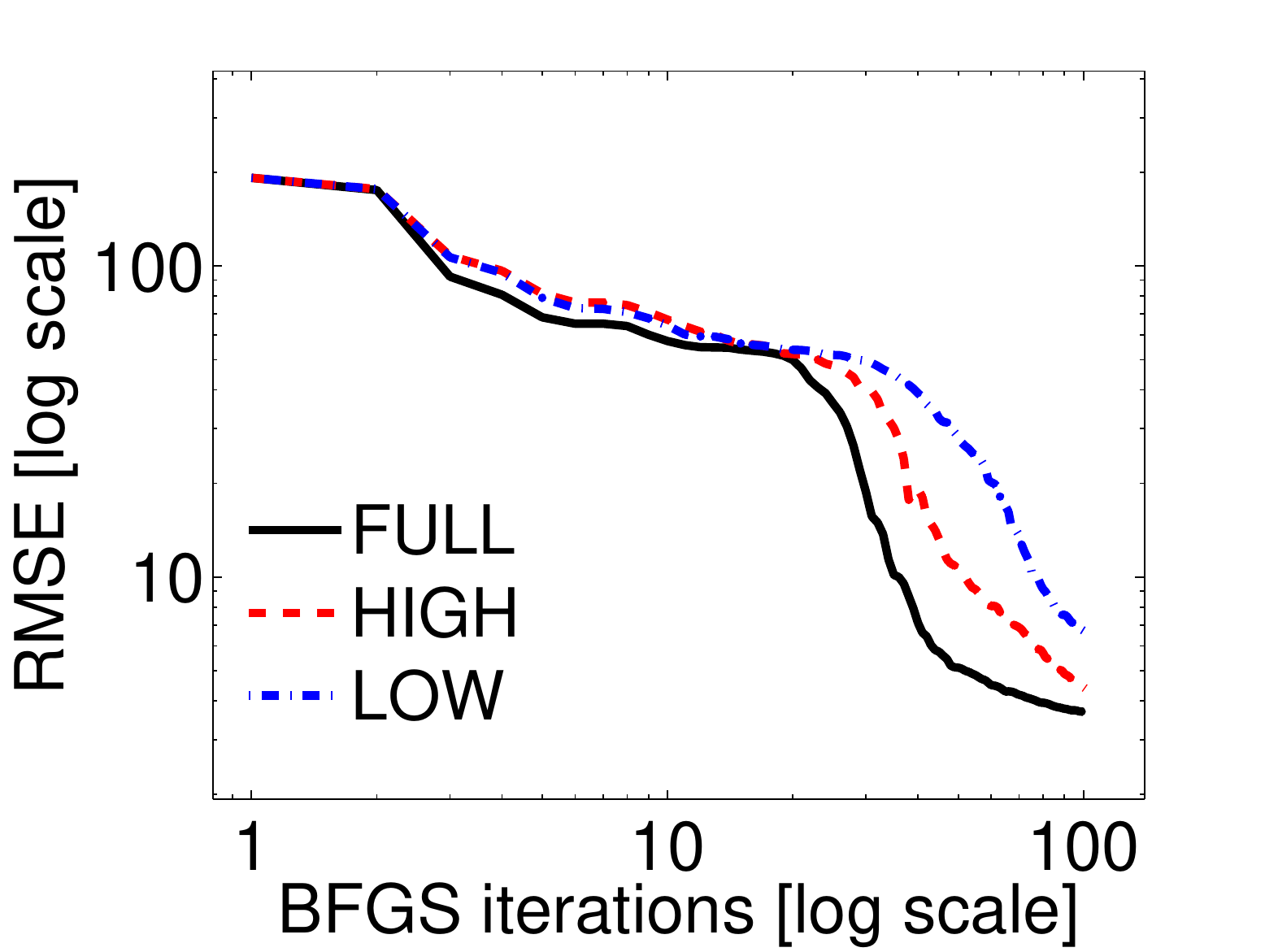} }
\subfigure[$u$ RMS error decrease versus the number of L-BFGS iterations.]{ \includegraphics[width=0.4\textwidth]{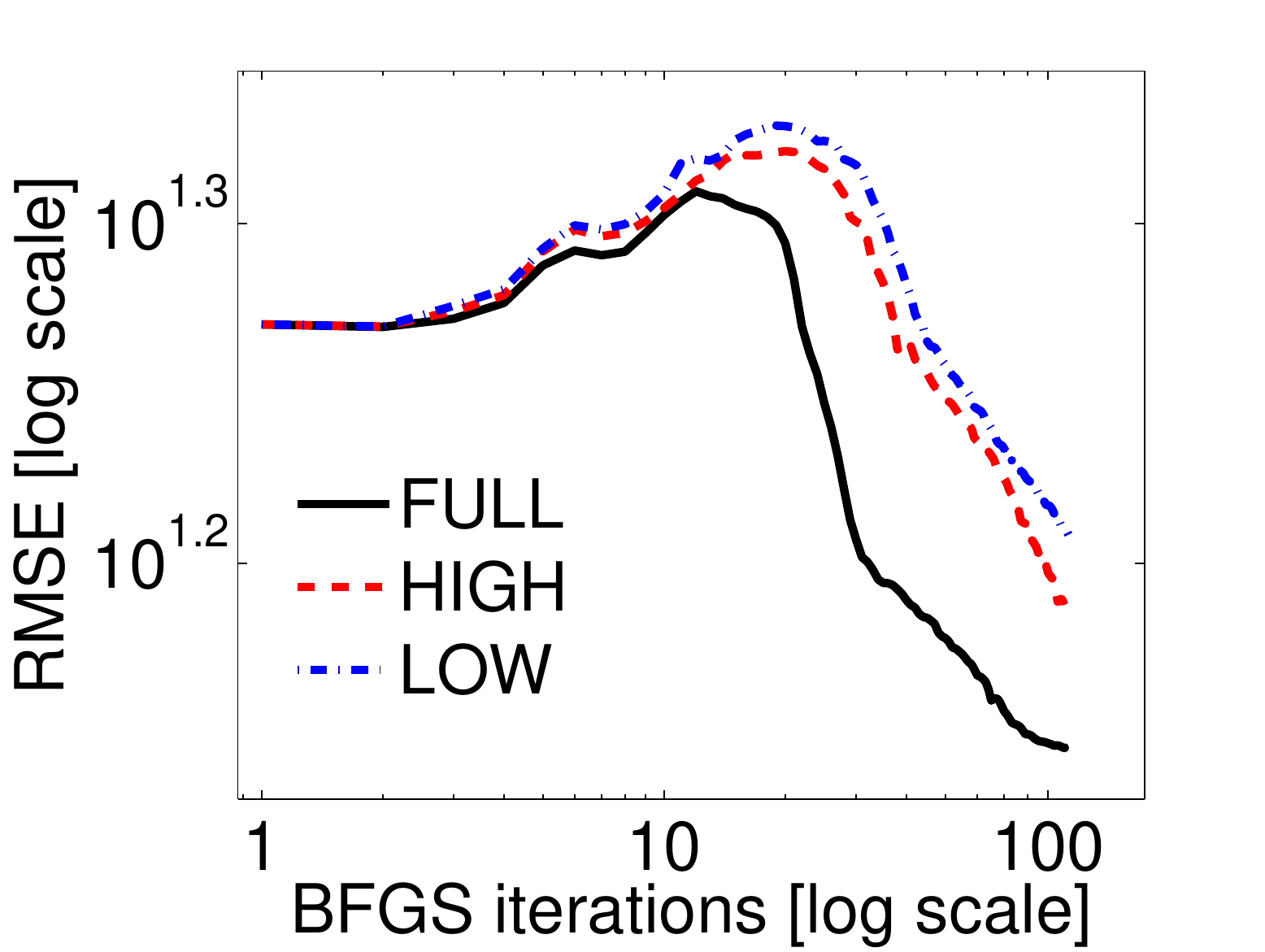} }
\subfigure[$v$ RMS error decrease versus the number of L-BFGS iterations.]{ \includegraphics[width=0.4\textwidth]{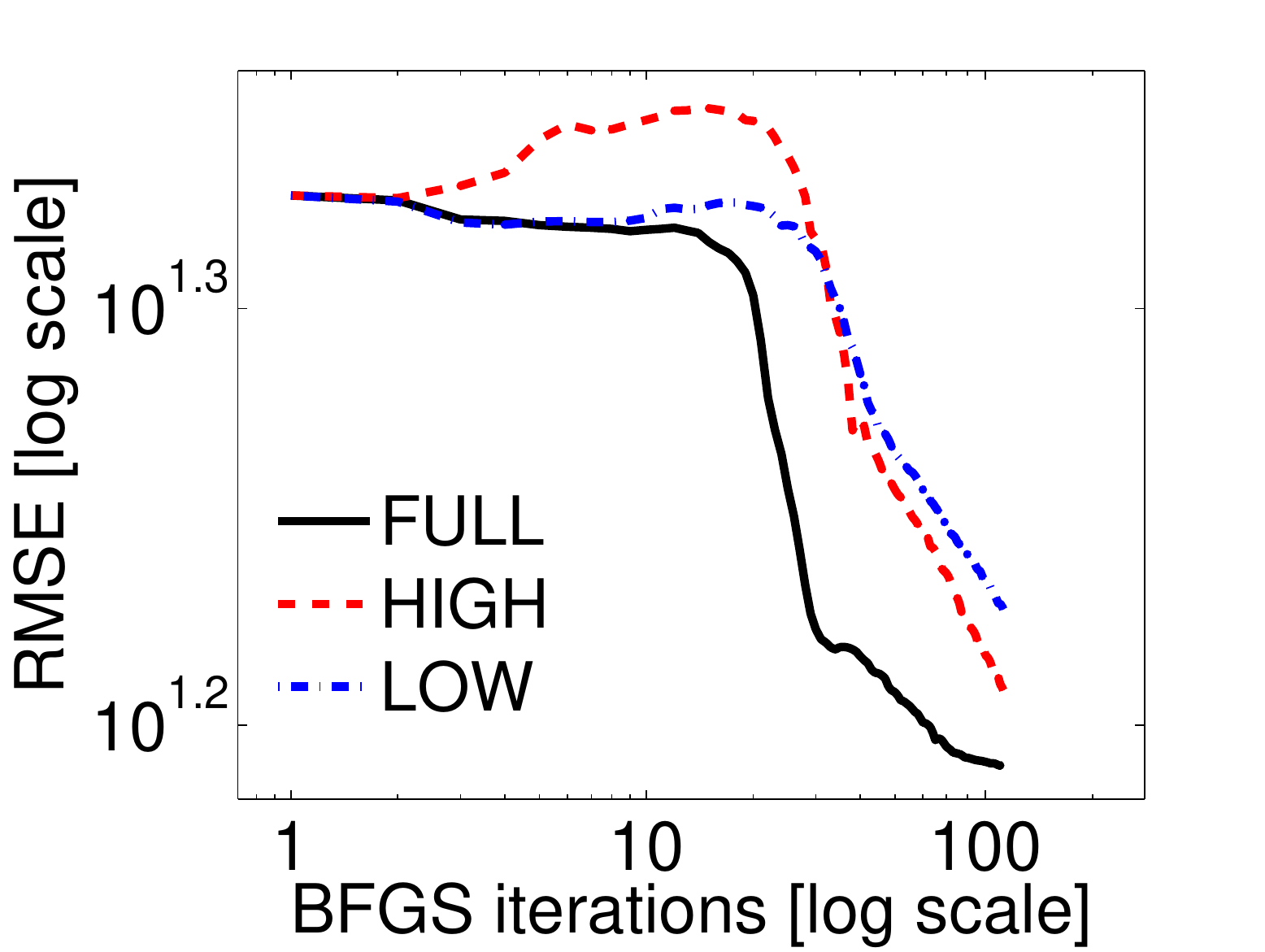} }
\subfigure[Location of {\sc high} (red) and {\sc low} impact observations]{ \includegraphics[width=0.4\textwidth]{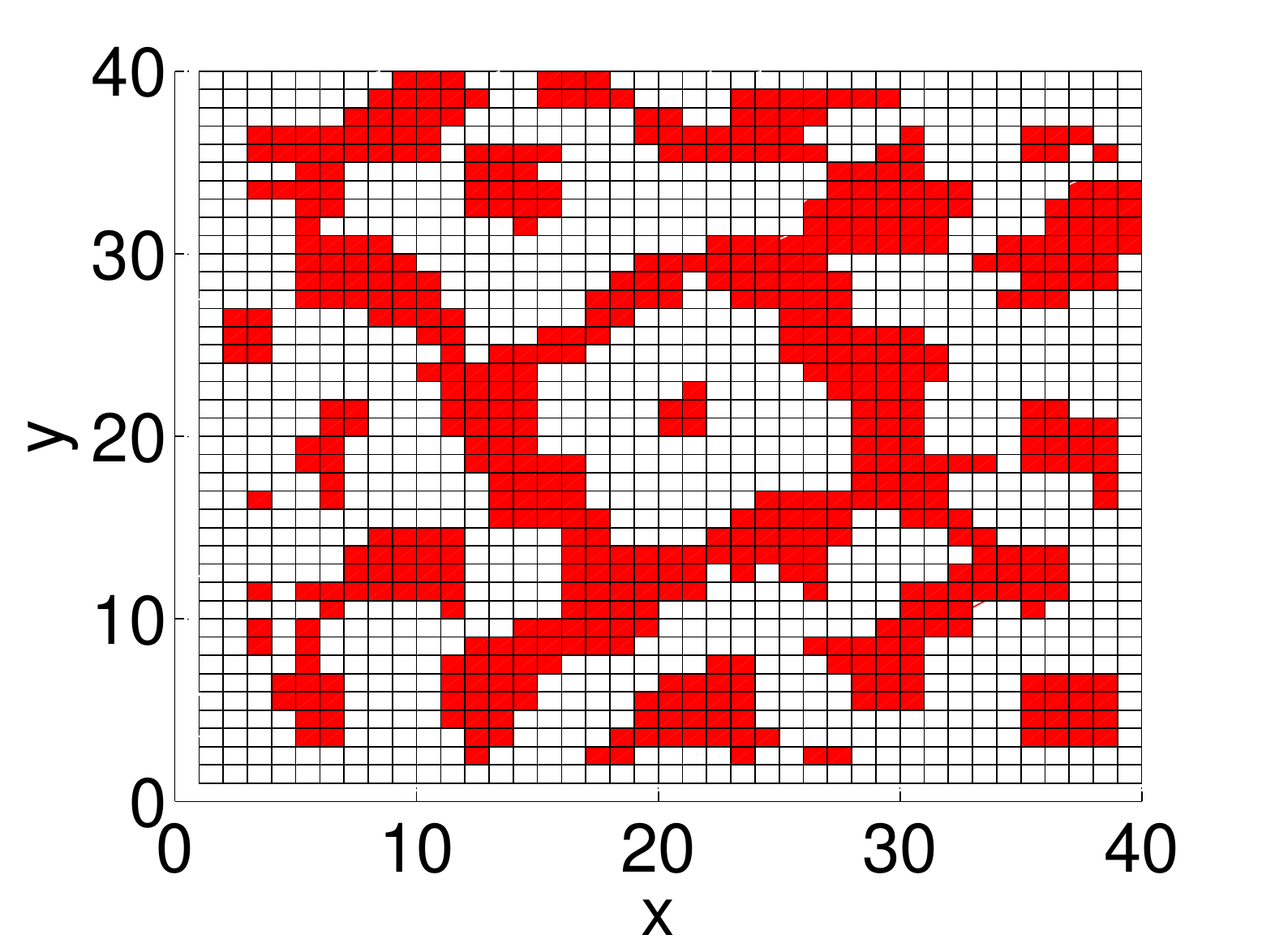} }
\caption{Location of high and low sensitivity observations, and their use in data assimilation.}
\label{fig:OI_mask}
\end{center}
\end{figure}

Figure \ref{fig:OI_mask}(a) shows the distribution of {\sc high} versus {\sc low} observations on our computational grid. 
We run 4D-Var to assimilate only the {\sc high}, followed by assimilating only the {\sc low} data points, 
and compare the decrease in the RMS true error in $h$ in Figure \ref{fig:OI_mask}(b). 
The convergence results show that assimilating observations of larger sensitivity yields slightly better results.
When the number of optimizer iterations increases, the two scenarios exhibit
similar performance, comparable to using all the observations.

%

\subsubsection{Identifying faulty data using observation sensitivity}

Finally, we consider the problem of detecting isolated faulty sensors and try to solve it using our new computational capabilities.
We prescribe certain sensors to exhibit the wrong behavior and expect the observation sensitivity analysis to reflect this.

We perform the 4D-Var scenario above again, using perfect observations at each grid point, with the exception of two points where
the observation values are increased by a factor of $10$ (thus simulating two faulty instruments). The two points are located at coordinates $(20,20)$, 
the grid center where fluid height field has large values initially, and at coordinates $(10,10)$, which is initially outside the water height profile.

\begin{figure}
\setcounter{subfigure}{0}
\centering
\subfigure[4D-Var increment]           {\includegraphics[width=0.4\textwidth]{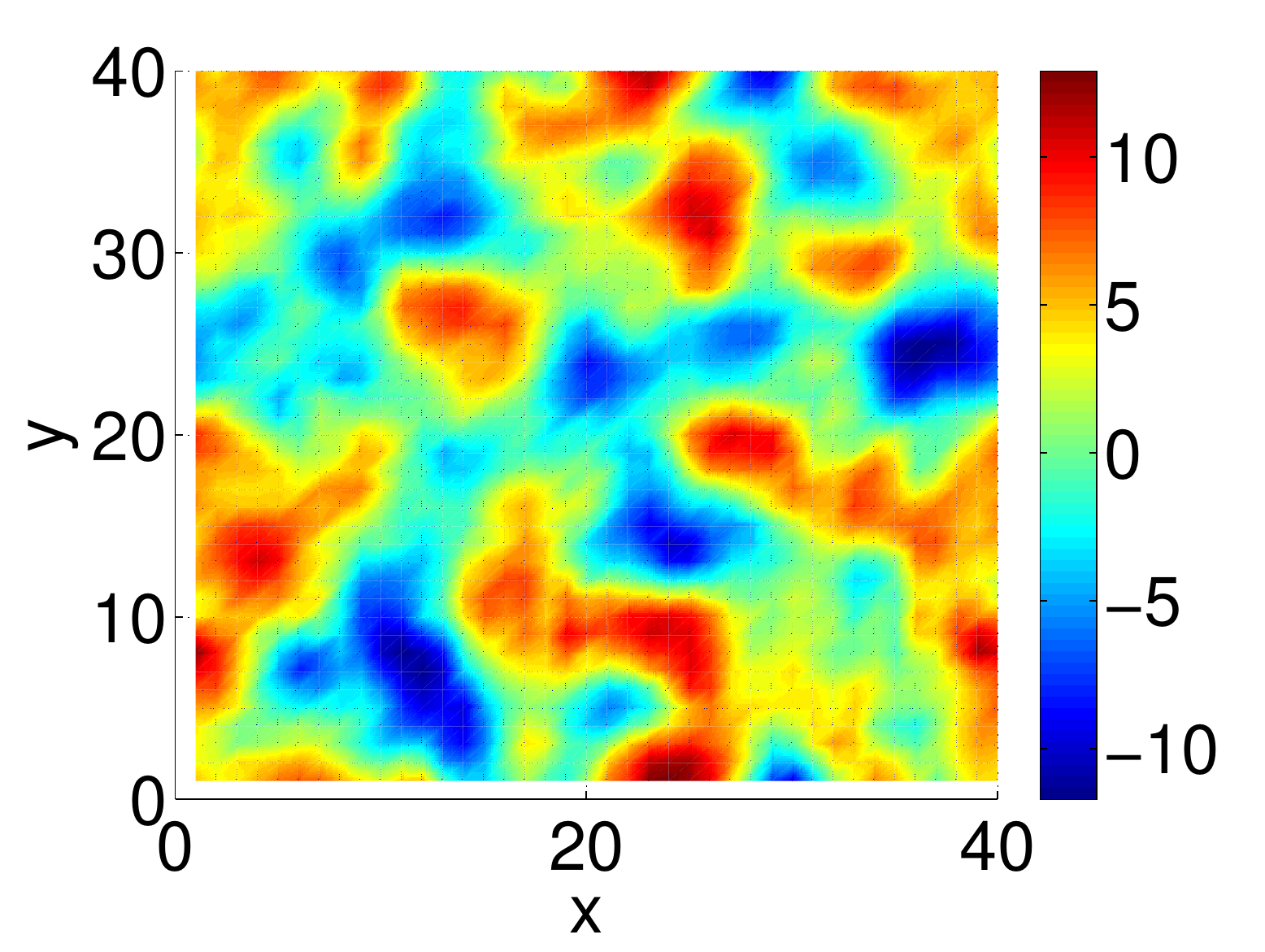}}
\subfigure[Supersensitivity field]     {\includegraphics[width=0.4\textwidth]{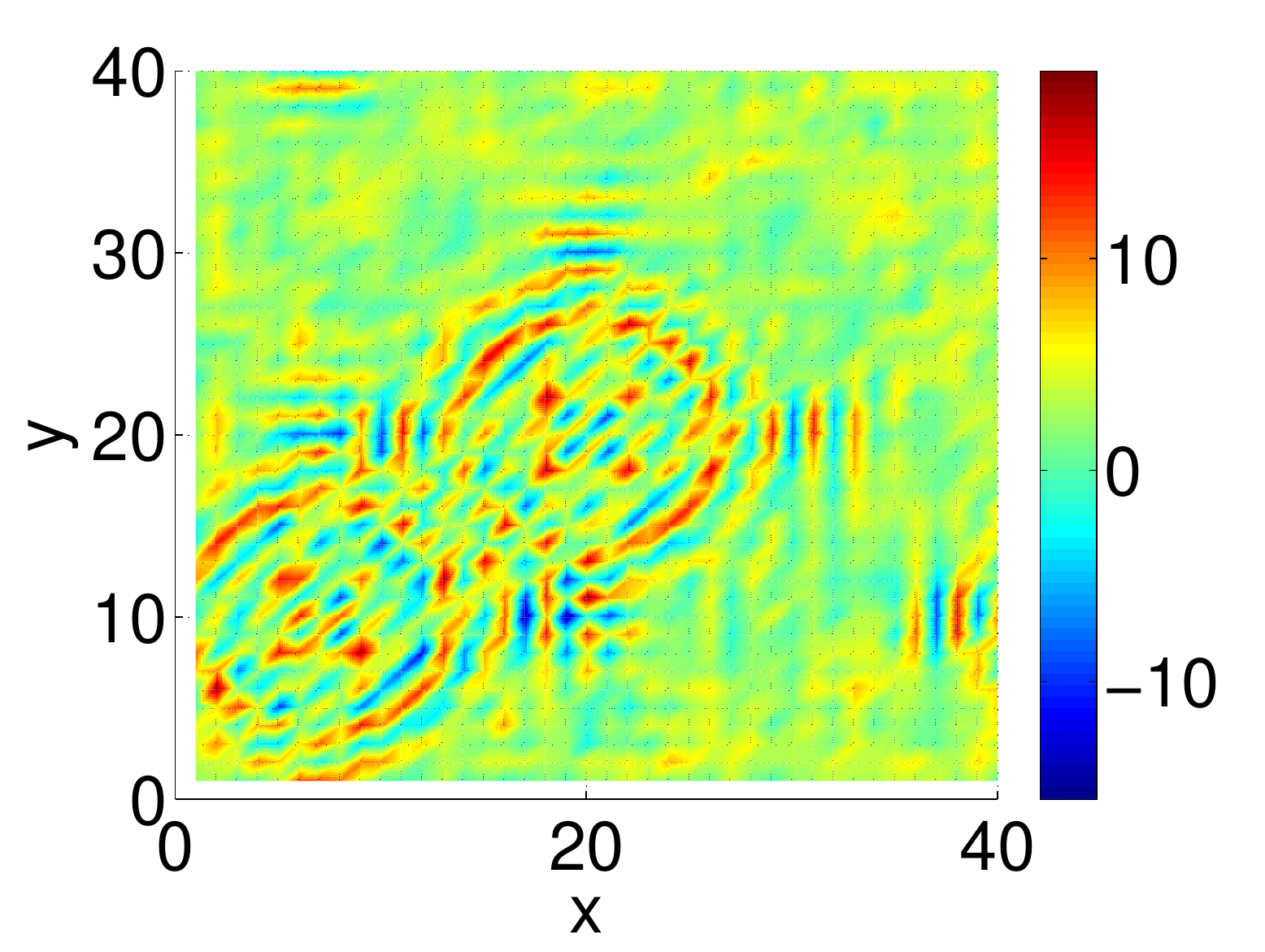}}
\subfigure[Sensitivity to observations]{\includegraphics[width=0.4\textwidth]{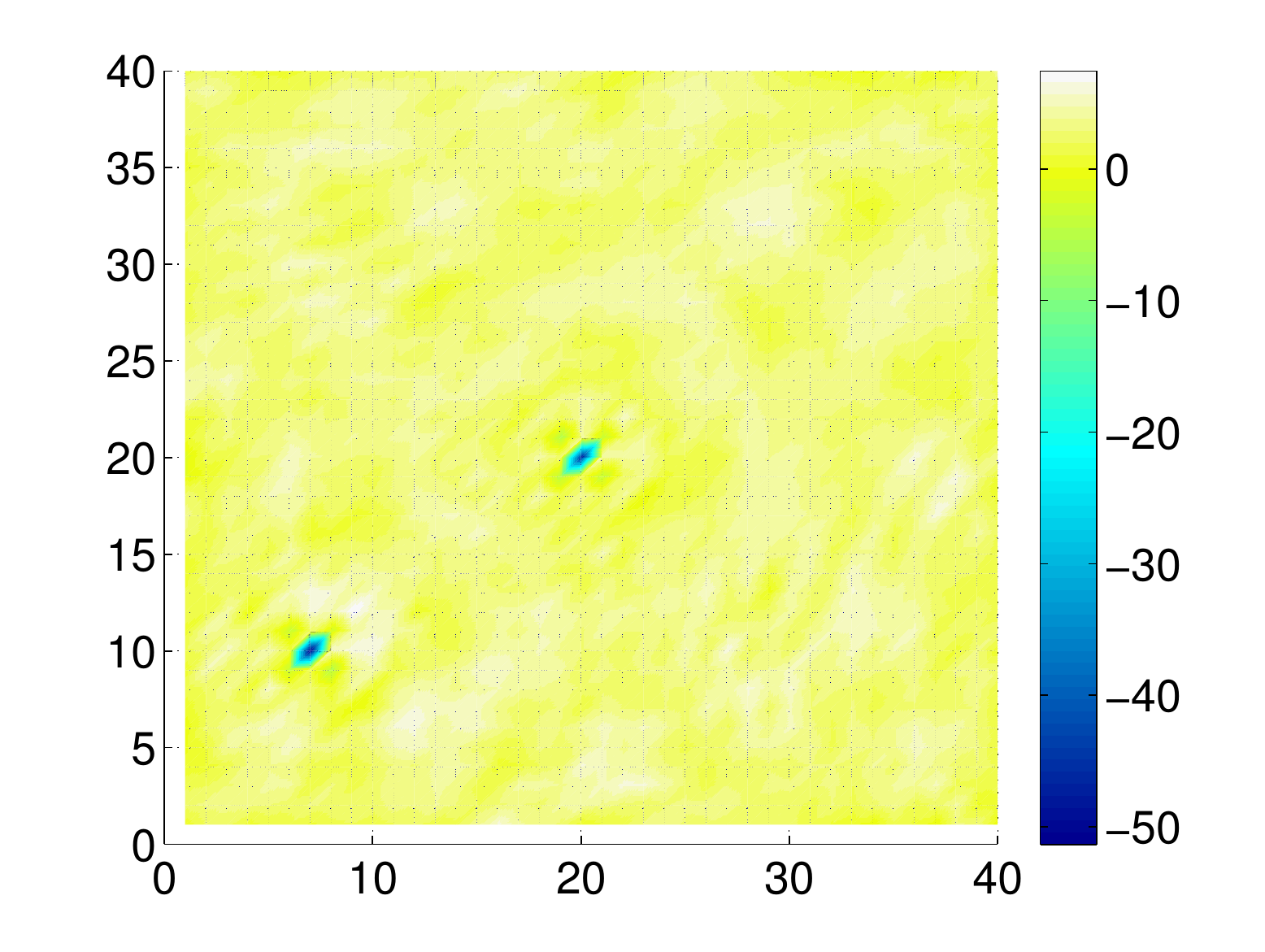}}
\caption{Observation sensitivity field when the assimilated data is corrupted at two locations with coordinates (10,10) and (20,20).
The location of the faulty sensors is unknown to the data assimilation system, but is retrieved via the observation impact methodology.}
\label{fig:faultysens}
\end{figure}

The 4D-Var data assimilation uses the faulty data together with the correct values.
Figure \ref{fig:faultysens}(a) plots the correction in the variable $h$, i.e., the quantity $(\xa_0 - \xb_0)$.
The correction field is smooth and there is no indication that there are very large errors in some data points.
Figure \ref{fig:faultysens}(b) plots the supersensitivity vector $\mu$ obtained by mapping the increment through the inverse 4D-Var Hessian (\ref{eqn:ST1_supersens}).
The supersensitivity field exhibits a clear structure consisting of large values in concentric circles around the two locations with faulty observations.
Figure \ref{fig:faultysens}(c) shows the observation sensitivity $\y$, obtained by propagating the supersensitivity through the tangent linear model (\ref{eqn:ST2_obssens}).
The observation impact methodology clearly identifies two points of abnormally large sensitivity. 
These are exactly the locations of the faulty sensors, 
and because their values are very different than the surrounding field, they have an abnormally large impact on the analysis.
This is precisely identified by the observation impact metrics.

\section{Conclusions and future work}\label{sec:concl}

This research is motivated by dynamic data driven applications which use measurements of 
the real system state to constrain computer model predictions. 
We develop efficient approaches to compute
the observation impact in 4D-Var data assimilation, i.e., the contribution of each data point to the final analysis.
Quantification of observation impact is important for applications such as data pruning, 
tuning the parameters of data assimilation,
identification of faulty sensors, and optimal configuration of sensor networks.
Immediate applications include  numerical weather predicition,
climate and air-quality forecast, hydrology, renewable energy systems and biohazard proliferation.

While 4D-Var is the state of the art data assimilation methodology, and is widely used in many operational centers, it is one of the most computationally challenging approaches to data assimilation. The computation of observation impact adds to the cost of performing analyses.
We review the derivation of the sensitivity equations that lead to the observation impact matrix,
and make in-depth comments about its structure, functionality and computation requirements.
One contribution of this work is to develop highly efficient implementation strategies for the
4D-Var sensitivity equations via a smart use of adjoint models. Two matrix free algorithms are proposed, one
serial and one parallel, to compute approximations of observation impact matrix.
These algorithms compute SVD-based low-rank approximations, which capture the most important features of observation impact matrix.
The accuracy of the generated approximations scales with the computational cost.

A second contribution of this work is to illustrate the use of observation impact 
in solving two practical applications: pruning observations of small impact, and detecting faulty sensors. 
Numerical experiments that validate the proposed computational framework are carried out with a two dimensional 
shallow water equations system.

Several future research directions emerge from this study.
On the computational side, the impact approximation algorithms can be further developed to achieve superior performance.
A rigorous analysis of approximation errors will be developed to guide the choice of the number of iterations and
the truncation level.
On the application side, our will be integrated in real large scale applications to provide a measure of importance 
of different measurements in real time assimilation.
In hindsight, the impact can be used to prune the data subsets and to detect erroneous data points.
In foresight, it can be used to design efficient strategies of sensor placement for targeted and adaptive observations.

\section*{Acknowledgements}

This work was supported by the National Science Foundation through the awards NSF DMS-0915047, NSF CCF-0635194, NSF CCF-0916493 and NSF OCI-0904397,
and by AFOSR DDDAS program through the awards FA9550--12--1--0293--DEF and AFOSR 12-2640-06.


\bibliographystyle{unsrt}
\bibliography{lowrank}

\end{document}